\documentclass{llncs}
\usepackage[ruled]{algorithm2e} 
\usepackage{amsmath}
\usepackage{esvect}
\usepackage{amsthm}
\usepackage{xcolor}

\title{New Characterizations of Strategy-Proofness under Single-Peakedness}

\author{Andrew B Jennings\inst{1} \and Rida Laraki\inst{2,3} \and Clemens Puppe\inst{4,5} \and Estelle Varloot\inst{2}}

\institute{Public Integrity Foundation
\and University of Liverpool \and Université Paris Dauphine-PSL \and Karlsruhe Institute of Technology \and Higher School of Economics}

\theoremstyle{plain}
\newtheorem{axiom}{Axiom}

\begin{document}

\maketitle
\begin{abstract}{We provide novel representations of strategy-proof voting rules when voters have uni-dimensional single-peaked preferences. In particular, we introduce a `grading curve' representation that is particularly useful in the variable electorate case. Our analysis recovers, links and unifies existing results in the literature, and provides new characterizations when strategy-proofness is combined with other desirable properties such as ordinality, participation, consistency, and proportionality. We also compare the computational properties of the various representations and show that the grading curve representation is superior in terms of computational complexity. Finally, the new representations are used to compute the strategy-proof methods that maximize the ex-ante social welfare for various $L_p$-norms and priors. For the $L_2$-norm and a uniform prior, the strategy-proof welfare maximizer is the linear median (or `uniform median'), that we also characterize as the unique proportional strategy-proof voting rule.}
\end{abstract}

\section{Introduction}
In mechanism design, strategy-proofness (SP) is a desirable property. It implies that whatever agents' beliefs are about others' behavior or information, their best strategy is to sincerely submit their privately-known types, even when their beliefs are wrong or mutually inconsistent. Consequently, strategy-proofness guarantees to the designer that she has implemented the intended choice function, i.e.~that the final decision is indeed linked in the intended way to the agents' true types. Of course, depending on the context, there are other desirable properties that one would like to satisfy, such as unanimity, voter sovereignty, efficiency, anonymity, neutrality, proportionality, and with variable electorate, consistency and participation (e.g. the absence of the no-show paradox).

When side payments are possible and utilities are quasi-linear, anonymous and efficient strategy-proof mechanisms can be designed (the well-known Vickrey-Clarke-Groves mechanisms). By contrast, in contexts of `pure' social choice (`voting'), the Gibbard-Satterthwaite \cite{Satterthwaite,Gibbard} Theorem shows that only dictatorial rules can be sovereign (`onto') and strategy-proof on an unrestricted domain of preferences. In particular, no onto voting rule can be anonymous and strategy-proof without restrictions on individual preferences.

To overcome the impossibility, several domain restrictions have been investigated. One of the most popular is one-dimensional single-peakedness. Under this domain restriction, the path-breaking paper by Moulin \cite{Moulin} showed that there is a large class of onto, anonymous and strategy-proof rules. All of them can be derived by simply adding some fixed ballots (called `phantom' votes) to the agents' ballots and electing the median alternative of the total. Moulin's paper inspired a large literature that obtained related characterizations for other particular domains or proved impossibility results (see, among many others, Border and Jordan \cite{Border/Jordan} and Barber\`a, Gul and Stacchetti \cite{Barbera}, Nehring and Puppe \cite{Nehring} and most recently, Freeman, Pennock, Peters and Wortman-Vaughan \cite{Freeman}).

\subsection*{Our Contribution}

In contrast to Moulin's elegant and simple phantom voter characterization in the anonymous case, the more general characterizations in terms of winning coalitions (called `generalized median voter schemes' in \cite{Barbera}, and `voting by issues' in \cite{Nehring}), as well as Moulin's
own `inf-sup' characterization in the appendix of his classic paper are complex.\footnote{Curiously,
Moulin provided completely different proofs for his two characterizations.} A basic objective of our paper is to provide simpler alternative representations that apply also to the non-anonymous case. Specifically, our main contributions are:

First, one of our representations is a natural extension of Moulin's idea: the selected outcome is the median of voters' peaks after the addition of new peaks computed from suitable \textit{phantom functions}. Another representation has a compact functional form which we refer to as a {\em grading curve} in the anonymous case. This is important as it allows to describe a family of voting rules with variable-sized electorate using one single function. This function may be interpreted as the density of the phantoms in Moulin's median formula.

Second, we compare the computational properties of the various representations and show that the grading curve representation is superior in terms of computational complexity. This is attractive in applications; moreover, the grading curve representation allows us to calculate the SP voting rule that maximizes the social welfare for various measures of the welfare.

Third, the new representations have additional merits. They can be used to connect the phantom characterization with the `voting by issues' one, and to provide new characterizations of and insights into special cases. For instance, we show that the case of uniformly distributed phantom voters (cf.~Caragiannis, Procaccia, and Shah \cite{ICML2016}) corresponds to a linear grading curve which in turn is fully characterized by one particular additional axiom: proportionality. We also show that this uniform ($=$ linear) median is the social utility maximizing strategy-proof voting rule when the utilitarian welfare is measured using the $L_2$-distance to the voters' peaks.

Fourth and finally, in the variable electorate environment, the curve representation allows us to tightly characterize all anonymous strategy-proof voting rules that are consistent in the sense of Smith \cite{Smith} and  Young \cite{young}: these are exactly those whose grading curve is independent of the size of the electorate.

\subsection*{Further Related Literature}

A link between Moulin's \cite{Moulin} inf-sup characterization and his phantom median voter characterization in the anonymous case has been provided by Weymark \cite{Weymark:2011} who showed how to derive the phantom median voter representation from the inf-sup one. Our paper shows that both results can be derived from our new characterization in terms of phantom functions, which also implies the one in terms of `winning coalitions' in \cite{Barbera} and \cite{Nehring}.

Freeman et al \cite{Freeman} prove that the median rule with uniformly distributed phantom voters is the unique anonymous, continuous and proportional strategy-proof method in a one-dimensional budget allocation problem if all voters have single peaked preferences. We show that anonymity is not necessary and, since we consider the proportionality axiom in a variable electorate context, we prove that continuity is also unnecessary in the characterization. Moreover, we show that the uniform median is the ex-ante utilitarian welfare maximizing SP rule when voters utilities are measured by the $L_2$-distance to the peaks and the ex-ante prior is given by the uniform distribution. A similar conclusion holds under an adversarial prior and the $L_2$-norm (which is the main result in Caragiannis et al \cite{ICML2016}; we extend this result to any $L_p$-norm, and also when voters have weights).

Extensions of the one-dimensional model analyzed here have been recently proposed
by Moulin \cite{Moulin2017} and Aziz et al \cite{Aziz2019}; the first paper offers a general framework that also covers the case of private (`rival') consumption, while the second paper considers the location
problem with capacity constraints.

\bigskip

The rest of the paper is organized as follows. Section 2 describes the model. Section 3 introduces our central phantom functions characterization that is used to reprove all the known representations of strategy-proof voting rules as well as several new representations. Section 4 considers additional properties, such as voter sovereignty and efficiency, strict responsiveness and ordinality, dummy voters, and anonymity. Variable electorate axioms such as consistency and participation (e.g.  absence of the no-show paradox) are considered in Section 5 where a complete characterization of consistent and/or participant methods is established in the anonymous and the non-anonymous cases.\footnote{To our knowledge, consistency has not yet been studied in the context of uni-dimensional strategy-proofness.} Section 6 computes the welfare maximizing voting rules under the strategy-proofness constraint. Section 7 concludes. The Appendix contains missing proofs and additional results (in particular, extensions to  voters with different weights, and to multi-dimensional separable and single-peaked preferences).

\section{Voting Model, Strategy-Proofness and its Consequences}

We start with a brief description of the voting model, and the introduction of the main property to be studied here: strategy-proofness.
The voting problem we are considering can be described by the following elements. First, an ordered set of alternatives $\Lambda$. For example political candidates on a left-right spectrum, a set of grades such as ``Great, Good, Average, Poor, Terrible'' or a set of locations on the line. Second, a finite set of voters $N=\{1,\dots,n\}$. A typical element of $ \vv{r} \in \Lambda^N$ is called a voting profile. Finally, a voting rule $\varphi$ is a function that associates to each voting profile in $\Lambda^N$ an element in $\Lambda$. 

This is interpreted as follows: each voter $i\in N$ has a single peaked preference over the linearly ordered set $\Lambda$ (see Definition \ref{single-peaked} below). He submits his peak (or a strategically chosen ballot) $r_i \in \Lambda$ to the designer who then computes $\varphi(r_1,..,r_n)=\varphi(\vv{r})$ and implements (or elects) the computed alternative. 

Without loss of generality we will assume $\Lambda \subseteq \mathbf{R}$ and use the notations $\mu^- := \mbox{inf } \Lambda$, $\mu^+ := \mbox{sup } \Lambda$ and $\overline{\Lambda} = \Lambda \cup \{\mu^-,\mu^+\}$. In Moulin's \cite{Moulin} paper, $\Lambda=\mathbf{R}$, $\mu^-=-\infty$ and $\mu^+=+\infty$. In Barber\`a, Gul and Stacchetti's \cite{Barbera} paper, $\Lambda$ is finite.

\begin{definition}[Single peaked preference] \label{single-peaked}
The preference order (e.g. the complete linear order) of voter $i$ over the alternatives in $\Lambda$ is single peaked if there is a unique alternative $x \in \Lambda$ such that for any $y, z \in \Lambda$, if $y$ is between $x$ and $z$, then voter $i$ prefers $x$ to $y$ and $y$ to $z$. The alternative $x$  is called the peak of the preference order. It is voter $i$'s favorite alternative.
\end{definition}

In terms of (ordinal) utility, this means that the utility function of each voter is increasing from $\mu^-$ to his peak and then decreasing from the peak to $\mu^+$.

\subsection{The Strategy-Proofness (SP) Axiom}

We wish the voting rule to satisfy some desirable properties. The main focus of this paper is strategy-proofness (SP). Sections 4 and 5 will explore various combinations with other axioms.

\begin{axiom}[Strategy-Proofness: SP]
A voting rule $\varphi$ is strategy-proof if for every voting profile $\vv{r}$ and voter $i\in N$, if $\vv{s}$ differs from $\vv{r}$ only in dimension $i$, then:

\[\varphi(\vv{s}) \geq \varphi(\vv{r}) \geq r_i\]

or 
\[\varphi(\vv{s}) \leq \varphi(\vv{r}) \leq r_i.\]
\end{axiom}

\begin{remark} The formulation of SP in Axiom 1 is usually called uncompromisingness \cite{Border/Jordan,EHLERS2002408}. It needs to be explained why it is analogous to the usual definition. The argument is as follows. If $r_i<\varphi(\vv{r})$  and $\varphi(\vv{s})<\varphi(\vv{r})$ (where $\vv{s}$ differs from $\vv{r}$ only in dimension $i$) then it is possible to create a single-peaked preference $P^{r_i}$ at $r_i$ such that $\varphi(\vv{s})$ is strictly preferred to $\varphi(\vv{r})$ by $P^{r_i}$. Hence, a voter with this preference, by reporting the peak $s_i$ instead of $r_i$, improves his utility, contradicting strategy-proofness. A similar conclusion is obtained for the other cases.
\end{remark}

\subsection{Consequences of the SP Axiom}

Here are some useful consequences of strategy-proofness.

\begin{definition}[Weak Responsiveness]
A voting rule $\varphi: \Lambda^{N} \rightarrow \Lambda$ is weakly responsive if for all voters $i$, and for all $\vv{r}$ and $\vv{s}$ that only differ in dimension $i$, if $r_i < s_i$ then $\varphi(\vv{r}) \leq \varphi(\vv{s})$.
\end{definition}

Weak responsiveness is sometimes called weak monotonicity. 

\begin{lemma}[Weak Responsiveness] \label{responsiveness}
If a voting rule $\varphi : \Lambda^N \rightarrow \Lambda$ is strategy-proof, then it is weakly responsive.
\end{lemma}

\begin{proof}
Let $\varphi$ be a voting rule that is SP. If $\vv{r}$ and $\vv{s}$ only differ in $i$ with $r_i < s_i$.
\begin{itemize}
    \item If $r_i < \varphi(\vv{r})$, then by strategy-proofness $\varphi(\vv{r}) \leq \varphi(\vv{s})$.
    \item If $s_i > \varphi(\vv{s})$, then by strategy-proofness $\varphi(\vv{r}) \leq \varphi(\vv{s})$.
    \item Otherwise, $\varphi(\vv{r}) \leq r_i < s_i \leq \varphi(\vv{s})$
\end{itemize}
\end{proof}

\begin{lemma} [Continuity] \label{continuity}
If a voting rule $\varphi : \Lambda^N \rightarrow \Lambda$ is strategy-proof, then it is continuous with respect to the $L_1$-norm on $\mathbf{R}$ (and so any norm on $\mathbf{R}$) .
\end{lemma}

\begin{proof}
Suppose that there is $\vv{r}$ and $\vv{s}$ that differ only in dimension $i$ such that $r_i < s_i$. First we show that $|\varphi(\vv{s}) - \varphi(\vv{r})| \leq s_i - r_i$. If $s_i < \varphi(\vv{s})$ or $r_i > \varphi(\vv{r})$, then strategy-proofness gives $\varphi(\vv{s}) = \varphi(\vv{r})$. Otherwise, $r_i \leq \varphi(\vv{r}) \leq \varphi(\vv{s}) \leq s_i$. In either case, $|\varphi(\vv{s}) - \varphi(\vv{r})| \leq s_i - r_i$.

Now let us use this property to show that $\varphi$ is continuous. Let $\epsilon >0$ be given. For any $\vv{r}$ and $\vv{s}$ with $|r_i - s_i |\leq \frac{\epsilon}{n}$ for all $i$, we have :
\begin{align*}
|\varphi(\vv{r}) - \varphi(\vv{s})| \leq & \sum_i |\varphi(r_1,\dots,r_i,s_{i+1},\dots,s_n) - \varphi(r_1,\dots,r_{i-1},s_i,\dots,s_n)| \\
\leq & \sum_i |r_i - s_i | \\
\leq & \epsilon.
\end{align*}
\end{proof}

\begin{lemma} [Continuous Extension]\label{continuous extension}
If a voting rule $\varphi : \Lambda^N \rightarrow \Lambda$ is strategy-proof, then it has a unique continuous extension in $\overline{\Lambda}^N \rightarrow \overline{\Lambda}$.
\end{lemma}

\begin{proof}
See Appendix \ref{Proof of Lemma continuous extension}.

\end{proof}

It is therefore natural to ask what are the SP voting rules in $\overline{\Lambda}^N \rightarrow \overline{\Lambda}$ that are not continuous extensions of voting rules in $\Lambda^N \rightarrow \Lambda$.

\begin{lemma}
A SP voting rule in $\overline{\Lambda}^N \rightarrow \overline{\Lambda}$ is not an extension of a voting rule in $\Lambda^N \rightarrow \Lambda$ iff it is constant valued with a value not in $\Lambda$.
\end{lemma}

\begin{proof}
$\Rightarrow:$ Suppose that $\varphi : \overline{\Lambda}^N \rightarrow \overline{\Lambda}$ is not an extension of a function from $\Lambda^N \rightarrow \Lambda$. Therefore, there is a voting profile $\vv{r} \in \Lambda^N$ such that $\varphi(\vv{r}) \not \in \Lambda$. Let $\varphi(\vv{r}) = \mu^-$ (resp. $\mu^+$). By strategy-proofness, for all $\vv{s} \in \overline{\Lambda}^N$,  $\varphi(\vv{s}) = \mu^-$ (resp. $\mu^+$). Therefore, $\varphi$ is a constant (equals to $\mu^-$ or to $\mu^+$) and its value is not in $\Lambda$.

$\Leftarrow:$ Immediate.
\end{proof}

From Lemma \ref{continuous extension}, by discarding the voting rules that are constant valued not in $\Lambda$, we obtain all the SP methods from $\Lambda$ to itself.
Consequently, from now on, we will consider without loss of generality that $\overline{\Lambda} = \Lambda$ (and so the voters are allowed to submit to the designer the extreme alternatives $\mu^-$ and $\mu^+$). 

\section{Characterizations of Strategy-Proof Voting Rules}

In this section we start by establishing two new and mathematically convenient characterizations of SP voting rules. The second of the two (more elegant) is a direct consequence of the first. In the subsequent sections, the second of the two is used to derive all the known characterizations as well as several new ones.

We denote by $\Gamma := \{\mu^-,\mu^+\}^N$ the set of voting profiles where voters have extreme positions (they submit an extreme alternative). For $X=(X_1,...,X_n)$ and $Y=(Y_1,...,Y_n)$ in $\Gamma$ we say that $X\leq Y$ if for every voter $i\in N$, $X_i \leq Y_i$.

\subsection{Phantom Function Characterizations}

In this subsection we will introduce the concept of phantom functions and two new characterizations of SP voting rules, the second being a direct consequence of the first. We show in the next sections that the second characterization implies not only all the known characterizations (the two by Moulin and the one by Barber\`a, Gul Stacchetti) but also several new representations. 

\begin{definition}[Phantom function]
A function $\alpha : \Gamma \rightarrow \Lambda$ is called a phantom function if $\alpha$ is weakly increasing ($X \leq Y \implies \alpha(X) \leq \alpha(Y)$). We will use the shorthand $\alpha_X := \alpha(X)$. 
\end{definition}

It is immediate that each SP voting rule $\varphi$ is associated with a unique phantom function $\alpha_{\varphi}$ defined as: 
\begin{equation}
\label{def_phantom}
    \forall X \in \Gamma, \alpha_{\varphi}(X) :=  \varphi(X).
\end{equation}  

That is, $\alpha_{\varphi}$ provides the outcome of $\varphi$ when all voters vote at the extremes. Observe that $\alpha_{\varphi}$ is necessarily weakly increasing because $\varphi$ is SP and so is weakly responsive (by Lemma \ref{responsiveness}). Conversely, the next theorem proves that each phantom function $\alpha$ is associated with a unique SP voting rule $\varphi_{\alpha}$. This is because strategy-proofness implies that we can always let voters vote at the extremes without changing the outcome. To state precisely our result, let us denote by $\theta$ that function which transforms voters' votes to the extreme alternatives.

\begin{definition}[The $\theta$ function]\footnote{The definition of $\theta$ is asymmetric because it sends values strictly below the cut-off to $\mu^-$ and greater than or equal to $\mu^+$. That's why our characterization in Theorem \ref{MCarac} ``looks'' asymmetric. It is not the case in the characterization of Theorem \ref{ICarac}.} \label{The theta function}
\begin{align*}
\theta : & \mathbf{\Lambda}^N \times \mathbf{R} \rightarrow \Gamma \\
\theta : & \vv{r},x \rightarrow X=\theta(\vv{r},x)
\end{align*}

Such that $\forall i \in N$, $r_i < x \implies ~X_i = \mu^- $ and $r_i \geq x \implies ~X_i = \mu^+ $
\end{definition}

Hence, $\theta$ at $x$ transforms each voter $i$ whose submitted input $r_i$ is strictly below (resp. above) $x$ to the extreme value $ \mu^-$ (resp. $\mu^+$). The next theorem shows that there is a one to one correspondence between SP rules and phantom functions, and provides a formula that relates them, thanks to the $\theta$ function.

\begin{theorem}[Phantom function characterization 1]\label{MCarac} The voting function $\varphi$ is strategy-proof iff there exists a phantom function $\alpha : \Gamma \rightarrow \Lambda$ such that:

\label{initial_characterization}

\begin{equation}
\forall \vv{r}\in \Lambda^n; \varphi(\vv{r}) := \left\{ \begin{array}{ll}
		\alpha_{\vv{\mu}^-} & \mbox{ if } \forall j, r_j \leq \alpha_{\vv{\mu}^-} \\
        \alpha_{\theta(\vv{r},r_i)} & \mbox{ if } (i \in N)
         \mbox{ and } r_i = min \{ r_j | r_j \geq \alpha_{\theta(\vv{r},r_i)} \} \\
        r_i & \mbox{ if }  (i\in N) 
        \mbox{ and } \forall \epsilon > 0, \alpha_{\theta(\vv{r},r_i + \epsilon)} \leq r_i \leq \alpha_{\theta(\vv{r},r_i)}    
    \end{array}
    \right.
\end{equation}

And that phantom function is necessarily unique.

\end{theorem}

\begin{proof}
The proof is quite long and so is delegated to Appendix \ref{proof of MCarac}. \end{proof}

As will be seen, Theorem \ref{MCarac} implies -- easily -- the next characterization which implies all the subsequent ones. The interpretation of Theorem \ref{MCarac} is simple. The first equation says that if all voters' peaks are smaller than the minimal value of the phantom function (which must coincide with the value of $\varphi$ when all voters vote $\mu^-$) then the outcome must be this value (thanks to weak responsiveness). The second case corresponds to the situation where the output is not one of the inputs ($\forall i\in N, \varphi(\vv{r}) \neq r_i$) and so there must exist an $i$ s.t. if all the voters whose peak is strictly below $r_i$ change their vote to $\mu^-$ and all the voters whose peak is above $r_i$ change their vote to $\mu^+$, then outcome of the voting rule does not change. The last case is when the output is one of the inputs. The $\epsilon$ that appears in this description shows why $r_i$ is a singular value. For $\epsilon>0$ small enough we have that $Y=\theta(\vv{r},r_i + \epsilon)$ differs from $X=\theta(\vv{r},r_i)$ only for voters $j$ such that $r_j=r_i$. Therefore $\theta(\vv{r},x)$ is discontinuous at $x=r_i$. Furthermore this discontinuity is impactful enough to cause $\alpha_Y \leq r_i \leq \alpha_X$. The use of the $\epsilon$ value is therefore practical as it allows us to describe $Y$ in terms of the $\theta$ function and show the relationship between $Y$, $X$ and $r_i$.

The next theorem provides a more elegant and compact characterization without the use of the $\theta$ function. For $X = (X_1, \dots, X_n) \in \Gamma$, we denote $\mu^-(X)$ (resp. $\mu^+(X)$) the set of voters  $i \in N$ such that $X_i = \mu^-$ (resp. $X_i= \mu^+$).

\begin{theorem}[Phantom function characterization 2]\label{ICarac} The voting function $\varphi$ is strategy-proof iff there exists a phantom function $\alpha : \Gamma \rightarrow \Lambda$ (the same as the one in theorem \ref{MCarac}) such that:
\begin{equation}
\forall \vv{r}; \varphi(\vv{r}) := \left\{ \begin{array}{lll}
		\alpha_X & \mbox{ if } \exists X \in \Gamma  \mbox{ s.t.} & \mu^+(X) \subseteq \{ j | \alpha_X \leq r_j \}\wedge \mu^-(X) \subseteq \{ j | \alpha_X \geq r_j \} \\
		& & \\
		r_i & \mbox{ if } \exists X, Y \in \Gamma  \mbox{ s.t.} &  \alpha_X \leq r_i \leq \alpha_Y \mbox{ and } \\ & & \mu^+(X) = \{ j | r_i < r_j \} \wedge \mu^-(Y) = \{ j | r_i > r_j \}
    \end{array}
    \right.
\end{equation}
\end{theorem}

\begin{proof}

Let $\varphi$ be defined as above and $\mu$ be defined as in the previous theorem:
\begin{equation*}
\forall \vv{r}; \mu(\vv{r}) := \left\{ \begin{array}{ll}
		\alpha_{\vv{\mu}^-} & \mbox{ if } \forall j, r_j \leq \alpha_{\vv{\mu}^-} \\
        \alpha_{\theta(\vv{r},r_i)} & \mbox{ if } (i \in N)
         \mbox{ and } r_i = min \{ r_j | r_j \geq \alpha_{\theta(\vv{r},r_i)} \} \\
       r_i & \mbox{ if }  (i\in N) 
        \mbox{ and } \forall \epsilon > 0, \alpha_{\theta(\vv{r},r_i + \epsilon)} \leq r_i \leq \alpha_{\theta(\vv{r},r_i)}    
    \end{array}
    \right.
\end{equation*}

Let us now compare the two.
\begin{itemize}
    \item If $\varphi(\vv{r}) = r_i$: Then for $X$ defined by $\mu^+(X) = \{j | r_i < r_j\}$ and $Y$ defined by $\mu^-(Y) = \{j | r_i > r_j \}$, we have $\alpha_X \leq r_i \leq \alpha_Y$. We also have $Y = \theta(\vv{r},r_i)$ and $X = \lim_{\epsilon >0 \wedge \epsilon \rightarrow 0} \theta(\vv{r},r_i + \epsilon)$. Therefore $\mu(\vv{r}) = r_i$
    \item Else $\varphi(\vv{r}) = \alpha_X$: We have $\mu^+(X) \subseteq \{ j | \alpha_X \leq r_j \}$ and $\mu^-(X) \subseteq \{ j | \alpha_X \geq r_j \}$. If $\mu^+(X) = \emptyset$ then $\mu(\vv{r})= \alpha_{\vv{\mu}^-}$ else there exists $r_i$ such that $r_i = \min\{r_j | r_j \geq \alpha_X\}$. It remains to show that $\alpha_{\theta(\vv{r},r_i)} = \alpha_X$. Let us suppose the opposite and reach a contradiction. Since $\mu^+(X) \subseteq \mu^+(\theta(\vv{r},r_i))$ and $X \neq \theta(\vv{r},r_i)$ there is $k \in \mu^+(\theta(\vv{r},r_i)) \cap \mu^-(X)$. Therefore $r_k \leq \alpha_X < \alpha_{\theta(\vv{r},r_i)} \leq r_k$. We have reached our contradiction. As such $\mu(\vv{r})=\alpha_X$.
\end{itemize}
It follows that $\mu=\varphi$.

\end{proof}

The intuition behind the characterization of Theorem \ref{ICarac} is simple. By strategy-proofness, when a voter's ballot is smaller than the societal outcome then, it can be replaced by the minimal ballot $\mu^-$ without changing the final outcome. Symmetrically, if it is larger than the final outcome, it can be replaced by $\mu^+$ without changing the outcome. As such, if the outcome is not one of the ballots, then it must be $\alpha_X$ where $\mu^+(X)$ is the set of voters whose ballots are greater than the outcome. On the other hand, if the outcome is one of the ballots then by weak responsiveness it is in between an $\alpha_X$ and $\alpha_Y$ where $\mu^+(X)$ is the set of voters whose ballots are  higher than the outcome and $\mu^-(Y)$ is the set of voters whose ballots are smaller than the outcome. 

\begin{remark}Observe that it is not possible to have both set inclusions in case 1 holding as equalities if the outcome is one of the peaks, for that would require
that anyone whose peak is the outcome would have both $\mu^-$ and $\mu^+$ associated with it, which is impossible.
\end{remark}

\begin{remark}The characterizations of $\varphi$ from $\alpha$ in Theorems \ref{MCarac} or \ref{ICarac} can be used to find $\varphi$ with an algorithm of complexity $O(2^n (n+f(n)))$ where $f(n)$ is the complexity of $\alpha$ for a given $X$ (Algorithm $\ref{algo:main}$ in the Appendix).
\end{remark}

\subsection{MaxMin Characterizations}

In his appendix, Moulin \cite{Moulin} proved the following MaxMin characterization of strategy-proof voting rules.

\begin{theorem}[Moulin's MaxMin Characterization]
\label{general_moulin}
A voting rule $\varphi$ is strategy-proof iff for each subset $S \subseteq N$ (including the empty set), there is a value $\beta_S \in \Lambda$ such that:

\begin{equation*}
\forall \vv{r}\in \Lambda^n, \varphi(\vv{r}) =\max_{S \subseteq N}  \min{(\beta_S, \inf_{i\in S}\{r_i\})}.
\end{equation*}
\end{theorem}

\begin{remark}
Moulin observed in his proof (the appendix in \cite{Moulin})  that without loss, $S \rightarrow \beta_S$ can be taken to be weakly increasing. With this selection, the $\beta$ in his theorem coincides with the $\alpha$ in Theorems \ref{MCarac} or \ref{ICarac} as proved now.
\end{remark}

\begin{theorem}[MaxMin Characterization 2] \label{corollary general_moulin}
A function $\varphi$ is strategy-proof iff there exists a phantom function $\alpha$ (the same as the one in Theorems \ref{MCarac} and \ref{ICarac}) that verifies:

\begin{equation}
\forall \vv{r}, \varphi(\vv{r}) =\max_{X \in \Gamma}  \min{(\alpha_X, \inf_{i\in \mu^+(X)}\{r_i\})}.
\end{equation}.
\end{theorem}

\begin{proof}

Let $\varphi$ be a strategy-proof voting rule defined by a phantom function $\alpha$ as in Theorem \ref{ICarac}. Let $\mu : \Lambda^{N} \rightarrow \Lambda$ be defined as in the theorem statement:

\begin{equation*}
\forall \vv{r}, \mu(\vv{r}) =\max_{X \in \Gamma}  \min{(\alpha_X, \inf_{i\in \mu^+(X)}\{r_i\})}.
\end{equation*}

Since $\varphi$ is the unique strategy-proof voting rule defined by $\alpha$ we only need to prove that $\varphi = \mu$.

Fix $\vv{r}$. Let $x = \mu(\vv{r})$ and choose $Z\in\Gamma$ that achieves the maximum, which means $\alpha_Z \geq x$ and $r_j \geq x$ for all $j\in\mu^+(Z)$. (Equivalently, $\mu^+(Z) \subseteq \{j|r_j \geq x\}$ or $\{j|r_j < x\} \subseteq \mu^-(Z)$.)

\begin{itemize}
    \item If $\mu(\vv{r}) = x = r_i$ for some $i$:
    
    Set $Y$ by $\mu^-(Y)=\{j|r_j < r_i\}$. Then $\mu^-(Y) \subseteq \mu^-(Z)$, so $\alpha_Y \geq \alpha_Z \geq r_i$.
    
    Set $X$ by $\mu^+(X)=\{j|r_j > r_i\}$. The maximum in the definition of $\mu(\vv{r})$ applies to $X$, so $\min (\alpha_X, \inf_{j\in\mu^+(X)}\{r_j\}) \leq r_i$. But $r_j > r_i$ for all $j\in\mu^+(X)$, which forces $\alpha_X \leq r_i$.
    
    This verifies that $\alpha_X \leq r_i \leq \alpha_Y$, so by the definition of $\varphi$, $\varphi(\vv{r}) = r_i = x$.
    
    \item If $\mu(\vv{r}) = x \neq r_i$ for any $i$: choose $X$ by $\mu^+(X) = \{j|r_j \geq x\} = \{j|r_j > x\}$ (since $x \neq r_i$ for any $i$). $\mu^+(Z) \subseteq \mu^+(X)$ so $\alpha_X \geq \alpha_Z \geq x$.
    
    The maximum in the definition of $\mu(\vv{r})$ applies to $X$, so \[\min(\alpha_X, \inf_{j\in\mu^+(X)}\{r_j\}) \leq x.\]
    But $r_j > x$ for $j\in\mu^+(X)$ which forces $\alpha_X \leq x$.
    
    It follows that $\alpha_X = x$ and we have constructed $X\in\Gamma$ such that $\mu^+(X)\subseteq\{j|\alpha_X\leq r_j\}\wedge\mu^-(X)\subseteq\{j|\alpha_X\geq r_j\}$. By the definition of $\varphi$, $\varphi(\vv{r}) = \alpha_X = x$.
\end{itemize}

Therefore, $\varphi = \mu$.
\end{proof}

\subsection{Median Characterizations}

The most popular of Moulin's representation (Theorem \ref{moulin_phantom}) assumes anonymity.

\begin{axiom}[Anonymity]
A voting rule $\phi$ is anonymous if for any permutation $\sigma$ and for all voting profiles $\vv{r}$:
\[\phi(r_{\sigma(1)},\dots,r_{\sigma(n)}) =\phi(r_1,\dots,r_n).\]
\end{axiom}

Anonymity states that all voters must be treated equally.

\begin{theorem}[Moulin's Median-Characterization. Anonymous Case] \label{moulin_phantom}
A voting rule $\varphi$ is strategy-proof and anonymous iff there is a set of $n+1$ values $\alpha_0,...,\alpha_{n}$ in $\Lambda$ (called phantom voters by Moulin) such that:
\[\forall \vv{r}; \varphi(\vv{r}) = med(r_1,\dots,r_n,\alpha_0,\dots,\alpha_{n}).\]
where $med$ denotes the median operator.\footnote{It is well-defined as we have an odd number of input values.} 
\end{theorem}

\begin{proof}
A direct consequence of Theorem \ref{Moulin-Type Characterization: General Case} and Proposition \ref{prop anonymity}.  
\end{proof}

The two characterizations of Moulin look quite different: median in the anonymous case (Theorem \ref{moulin_phantom}) and maxmin in the general case (Theorem \ref{general_moulin}). 

Their proofs are separated in his article. In order to link the two we need to be able to choose $n+1$ phantom voters among the $2^N$ outputs of our phantom functions. The next theorem explains how they can be chosen.

For all $k=0,...,n$ and $\vv{r}= (r_1,\dots,r_n)$, let $X_k(\vv{r}) \in \Gamma$ be defined in such a way that $\mu^+(X_k(\vv{r}))$ is equal to the set of voters that provides the $k$ largest peaks.\footnote{If there are more than $k$ voters with the $k$ largest peaks, break ties arbitrarily. The tie breaking rule does not affect the outcome.} For $k=0$, we let $X_0(\vv{r})=\vv{\mu^-}=(\mu^-,...,\mu^-).$

\begin{theorem}[Median Characterization. General Case]\label{Moulin-Type Characterization: General Case}
A voting rule $\varphi$ is strategy-proof iff there exists a phantom function $\alpha$ (the same as in Theorem \ref{ICarac}) such that:

\begin{equation}
    \forall \vv{r}; \varphi(\vv{r}) := med(r_1,\dots,r_n,\alpha_{X_0(\vv{r})},\alpha_{X_1(\vv{r})},\dots,\alpha_{X_n(\vv{r})}).
\end{equation}

\end{theorem}

\begin{proof}

This is a direct consequence of Theorem \ref{ICarac}. See Appendix \ref{Proof of Theorem Median General}.
\end{proof}

\begin{remark}
This characterization implies easily the Moulin's median in Theorem \ref{moulin_phantom} (thanks to Proposition \ref{prop anonymity} below). Hence, we have unified the proofs of the two Moulin's characterizations. Weymark \cite{Weymark:2011} unified the two, using a different approach.

\end{remark}

\begin{remark}
The last characterization can be used to compute $\varphi(\vv{r})$ with the complexity $O(n\log(n)+nf(n))$ (Algorithm $\ref{algo:moulin}$ in the Appendix). This is better than the complexity of the phantom characterization ($O(2^n (n+f(n)))$).
\end{remark}

\subsection{Curve Characterizations}

This section establishes two novel characterizations of SP methods.

\begin{theorem}[Curve Characterization. General Case]\label{theo curve}
A voting rule $\varphi$ is strategy-proof iff there exists a phantom function $\alpha$ (the same as the one in Theorems \ref{MCarac} and \ref{ICarac}) such that:

\[\forall \vv{r}; \varphi(\vv{r}) := \sup \left\{ y \in \Lambda | \alpha_{\theta(\vv{r},y)} \geq y \right\}.\]
Where the $\theta$ function is as in Definition \ref{The theta function}.
\end{theorem}

\begin{proof}
Again, it is a direct consequence of Theorem \ref{ICarac}. Let $\varphi$ be a strategy-proof voting rule defined by a phantom function $\alpha$. And let $\mu$ be defined as 
\[\forall \vv{r}; \mu(\vv{r}) := sup \left\{ y \in \Lambda | \alpha_{\theta(\vv{r},y)} \geq y \right\}.\]

Since $\varphi$ is the unique strategy-proof voting rule defined by $\alpha$ we only need to prove that $\varphi = \mu$.

First notice that $y \rightarrow \alpha_{\theta(\vv{r},y)}$ is non-increasing.

\begin{itemize}
    \item Case $\varphi(\vv{r}) = \alpha_X$: By the characterization of $\varphi$, we have $\mu^+(X) \subseteq \{j|\alpha_X \leq r_j\}$ and $\mu^-(X) \subseteq \{j | \alpha_X \geq r_j\}$.
    
    The former implies that $\alpha_X \leq \alpha_{\theta(\vv{r}, \alpha_X)}$.
    
    The latter implies that $\{j | \alpha_X < r_j\} \subseteq \mu^+(X)$. For $\epsilon > 0$, we have \[\{j | \alpha_X + \epsilon \leq r_j\} \subseteq \{j | \alpha_X < r_j\} \subseteq \mu^+(X)\] and $\alpha_{\theta(\vv{r}, \alpha_X + \epsilon)} \leq \alpha_X < \alpha_X + \epsilon$.
    
    Therefore
    $\{y \in \Lambda | \alpha_{\theta(\vv{r},y)}\geq y\}$
    must contain $\alpha_X$ and not $\alpha_X + \epsilon$.
    It follows that $\mu(\vv{r}) = \alpha_X$.
    \item Case $\varphi(\vv{r}) = r_i$: Let $X$ and $Y$ be such that $\alpha_X \leq r_i \leq \alpha_Y$ and $\mu^+(X) = \{ j | r_i < r_j \} \wedge \mu^-(Y) = \{ j | r_i > r_j \}$. For any $\epsilon > 0$ we have:
    \[\alpha_{\theta(\vv{r},r_i + \epsilon)} \leq \alpha_X \leq r_i \leq \alpha_Y = \alpha_{\theta(\vv{r},r_i)}.\]
    Therefore $\alpha_{\theta(\vv{r},r_i + \epsilon)} \leq r_i < r_i + \epsilon$ and $r_i \leq \alpha_{\theta(\vv{r},r_i)}$. Thus 
    $\{y \in \Lambda | \alpha_{\theta(\vv{r},y)} \geq y\}$
    must contain $r_i$ and not $r_i + \epsilon$. As such $\mu(\vv{r}) = r_i$.
\end{itemize}

Therefore $\mu = \varphi$.
\end{proof}

\begin{remark} This representation yields Algorithm \ref{algo:jennings} (see appendix) that returns $\varphi(\vv{r})$ with a computational complexity $O(log(n)(n + f(n)))$. It improves upon the previous characterizations.
\end{remark}

\begin{remark}  The curve characterization is not only less complex than the other characterizations, but it is also useful in the proof and/or characterization of dummy voters (Section 4.3), participation (Section 5.1), consistency (Sections 5.2 and 5.3), proportionality (Section 5.4), and social welfare maximization (Section 6). In the anonymous case, the characterization can be simplified as follows.
\end{remark}

The grading curve representation in the next result is one of the central contributions of our paper.

\begin{theorem}[Curve Characterization. Anonymous Case]\footnote{The result first appeared in the unpublished PhD thesis \cite{Jennings} of our co-author A.~Jennings.} \label{functional anymous characterization}
A voting rule $\varphi : [0,R]^n \rightarrow [0,R]$ is strategy-proof and anonymous iff there is an increasing function $g:[0,R] \rightarrow [0,1]$ with $g(y) > 0$ for $y > 0$ such that:
\[\forall \vv{r}, \varphi(\vv{r}) := \sup \left\{ y | \dfrac{\#\{ r_i \geq y\}}{n} \geq g(y) \right\}.\]
The $g$ function is called the grading curve associated to $\varphi$.
\end{theorem}

\begin{proof}
A direct consequence of Theorem \ref{theo curve} above and Theorem \ref{lemma-Curve} below.  
\end{proof}

The $g$ function has a very intuitive interpretation as the density of the phantom voters. In fact, the median representation in Theorem \ref{moulin_phantom} needs the specification of $n+1$ values, and those values change completely with the size $n$ of the electorate. By contrast, under the curve representation, the same function can describe a family of mechanisms for all $n$ simultaneously. For example, the linear function $g(x)=\frac{x}{R}$ corresponds to the $n+1$ phantoms uniformly distributed across the interval $[0,R]$, for all jury size $n$ (see Section 5 below for more details).


\subsection{Voting by Issues Characterizations}

The `voting by issues' representation established by Barber\`a, Gul and Stacchetti \cite{Barbera} appears less practical than the other formulations, but it has the great power of being extendable to multi-dimensional and generalized single-peaked domains (see Nehring and Puppe \cite{Puppe,Nehring}). In this section, we state the original result and then refine it using our new tools. 

\begin{axiom}[Voter Sovereignty]\label{Voter sovereignty}
A voting rule $\phi$ is voter sovereign if for all $x \in \Lambda$ there is a preference profile $\vv{r}$ such that:
$\phi(\vv{r}) = x$.
\end{axiom}

This means that all alternatives can potentially be selected as the outcome. Section 4.1 characterizes voter sovereignty in terms of phantom functions. 

A \emph{voting by issues} consists of a property space $(\Omega,\mathcal{H})$ where $\Omega$ is a set of alternatives and $\mathcal{H}$ is a set of subsets of $\Omega$ called \emph{properties}. Each property $H$ comes together with its complementary property $H^c := \Omega \setminus H$; the pair $(H,H^c)$ is called an \emph{issue}. Each voter $i$ provides a ballot $r_i \in \Omega$. Each issue $(H,H^c)$ is then resolved separably as a binary election. If $r_i \in H$ then we say that voter $i$ votes for issue $H$ or that his ballot verifies issue $H$. And if $r_i \not \in H$ then we say that he voted against $H$, i.e.~he voted for $H^c$. If an issue is elected then the result of the vote will be one of the elements of the issue. As such, the result of the election is an intersection of elements of $\mathcal{H}$. A coalition is a subset of voters. A coalition $W \subseteq N$ is said to be winning for $H\in \mathcal{H}$, if when all voters in $W$ voted for $H$, $H$ is elected. Let $W_H$ be the set of winning coalitions for $H$. The following result is due to Barber\`a et al. \cite{Barbera}; our present formulation follows \cite{Nehring}.

\begin{theorem}[Voting by Issues Characterization] Let $\Lambda$ be finite. A voting rule $\varphi:\Lambda^{N} \rightarrow \Lambda$ is strategy-proof and voter-sovereign iff it is voting by issues satisfying, for all $G, H \in N, G \subseteq H \Rightarrow \mathcal{W}_G \subseteq \mathcal{W}_H$. 
\end{theorem}

This result was proved with $\Lambda$ finite, assuming voter-sovereignty. In the next characterization we explicitly define the winning coalitions in terms of the phantom functions without assuming voter sovereignty nor that $\Lambda$ is finite.

\begin{theorem}[Explicit Voting by Issues Characterization] \label{An Explicit Voting by Issue Characterization}
A voting rule $\varphi$ is strategy-proof iff it is voting by issues on the property space $(\Lambda,\mathcal{H})$ where $\mathcal{H}$ consists of all properties of the form
$\{x\in \Lambda:x\leq a\}$, $\{x\in \Lambda:x\geq a\}$, and their complements, and such that, for all $a\in \Lambda$:
\begin{itemize}
\item $\mu^+(X)$ is a winning coalition for $H=\{y \in \Lambda: y \geq a\}$ if and only if $\alpha_X \geq a$;
\item $\mu^-(X)$ is a winning coalition for $H=\{y \in \Lambda: y \leq a\}$ if and only if $\alpha_{X} \leq a$.
\end{itemize}
The phantom function $\alpha$ is the one associated to $\varphi$, as in Theorems \ref{MCarac} and \ref{ICarac}.

\end{theorem}

\begin{proof}
Let $\varphi$ be a strategy-proof voting rule defined by a phantom function $\alpha$ as in Theorem \ref{ICarac}. Let $\mu$ be the vote by issue given in Theorem \ref{An Explicit Voting by Issue Characterization}.

Since $\varphi$ is the unique strategy-proof voting rule defined by $\alpha$ we only need to prove that $\varphi = \mu$.

\begin{itemize}
    \item Case $\varphi(\vv{r}) = \alpha_X:$ For $a = \alpha_X$ we have $\mu^+(X) \subseteq \{ j | r_j \geq a\}$ is a winning coalition for $\{ y \geq a\}$ and $\mu^-(X) \subseteq \{j | r_j \leq a\}$ is a winning coalition for $\{ y \leq a\}$. Therefore $\mu(\vv{r}) = a$.
    \item Case $\varphi(\vv{r}) = r_i:$ Let $X, Y \in \Gamma$ be the same voting profiles as found in Theorem \ref{ICarac}. For $a = r_i$ we have $\alpha_Y \geq a$ therefore $\mu^+(Y) \subseteq \{j | r_j \geq a\} $ is a winning coalition for $\{ y \geq a\}$ and $\alpha_X \leq a$ therefore $\mu^-(X) \subseteq \{ j | r_j \leq a\}$ is a winning coalition for $\{ y \leq a\}$. Therefore $\mu(\vv{r}) = a$.

\end{itemize}

We have shown that $\varphi = \mu.$

\end{proof}

\subsection{Computational Complexity}

The following table recalls the different characterizations and their complexity. The value $f(n)$ is the time complexity of $\alpha$ when the input is of size $n$.
{\small
\begin{center}\label{table charac}
    \begin{tabular}{|l|l|l|}
    \hline
         Characterization & Formula & Complexity\\
         \hline
        Phantom Function & $\varphi(\vv{r}) := \left\{ \begin{array}{lll}
		\alpha_X & \mbox{ if } &\forall j, \mu^+(X) = \{ j | \alpha_X \leq r_j \}\wedge \mu^-(X) = \{ j | \alpha_X \geq r_j \} \\
		& & \\
		r_i & \mbox{ if } &  \alpha_X \leq r_i \leq \alpha_Y \mbox{ where } \\ & & \mu^+(X) = \{ j | r_i < r_j \} \wedge \mu^-(Y) = \{ j | r_i > r_j \}
    \end{array}
    \right. $ & $2^n (n+f(n))$ \\
        \hline
        & & \\
        MaxMin & $\varphi(\vv{r}) =\max_{X \in \Gamma}  \min{(\alpha_X, \inf_{i\in \mu^+(X)} \{ r_i\})}$ & $2^n (n+f(n))$\\
        & & \\
        \hline
        Median & & \\
        & $\varphi(\vv{r}) := med(r_1,\dots,r_n,\alpha_{X_0(\vv{r})},\alpha_{X_1(\vv{r})},\dots,\alpha_{X_n(\vv{r})})$ & $n(log(n) + f(n))$\\
       & & \\
        \hline 
        Curve & & \\
         & $\varphi(\vv{r}) := \mbox{sup } \{y | \alpha_{\theta(\vv{r},y)} \geq y \}$ & $log(n)(n + f(n))$\\
        & & \\
        \hline
        & & \\
        Voting by issues & $\varphi(\vv{r}) := a$  :  $\{r_i \leq a\} \in \{\mu^+(X) | \alpha_X \geq a\} \wedge \{r_i \geq a\} \in \{\mu^-(X) | \alpha_X \leq a\}$ & $2^n (n+f(n))$\\ 
        &  &  \\
        \hline
    \end{tabular}
\end{center}

}

\section{Additional Properties: Fixed Electorate}

The phantom function and the different representations (in particular the curve representation) will be very helpful in understanding the effects of imposing more axioms on the voting rule. This is the subject of this and the next section.

\subsection{Voter Sovereignty and Efficiency}

In many applications, it makes little sense to vote for an alternative that is never a possible output. Therefore, one may wish the voting rule to be voter-sovereign, that is for all $x \in \Lambda$ there is a preference profile $\vv{r}$ such that:
$\phi(\vv{r}) = x$ (see Axiom \ref{Voter sovereignty} above).

\begin{proposition}
A strategy-proof voting rule $\varphi : \Lambda^{N} \rightarrow \Lambda$ is voter-sovereign iff its phantom function $\alpha$ satisfies $\alpha_{\vv{\mu}^-} = \mu^-$ and $\alpha_{\vv{\mu}^+} = \mu^+$. In that case, $\varphi$ is unanimous ($\varphi(a,\dots,a) = a$ for all $a\in \Lambda$).
\end{proposition}

\begin{proof}
$\Rightarrow$ If $\alpha_{\vv{\mu}^-} > \mu^-$ (resp. $\alpha_{\vv{\mu}^+} < \mu^+$), then by our characterization, (for example Theorem \ref{Moulin-Type Characterization: General Case}), we have $\varphi(\vv{\mu}^-) = \alpha_{\vv{\mu}^-}$.
Let $x$ be any value such that $x < \alpha_{\vv{\mu}^-}$. By weak responsiveness (Lemma \ref{responsiveness}), we have $ \forall \vv{r}, \varphi(\vv{r}) \geq \alpha_{\vv{\mu}^-} > x$. As such, $\varphi$ is not voter-sovereign since $x$ cannot be reached.

$\Leftarrow$ Suppose that  $\alpha_{\vv{\mu}^-} = \mu^-$ and $\alpha_{\vv{\mu}^+} = \mu^+$. Then $\varphi(a,\dots,a) = a$ for all $a\in \Lambda$ (use for example the characterization in Theorem \ref{Moulin-Type Characterization: General Case} to deduce it). As such we are voter-sovereign.
\end{proof}

\begin{axiom}[Pareto Optimality]
A voting rule is called Pareto optimal if no different alternative could lead to improved satisfaction for some voter without loss for any other voter.
\end{axiom}

\begin{proposition}
If a voting rule $\varphi : \Lambda^{N} \rightarrow \Lambda$ is voter-sovereign and strategy-proof then it is efficient (the selected alternative is Pareto optimal).
\end{proposition}

This result was proved in Weymark \cite{Weymark:2011}. Here is an alternative proof.

\begin{proof}
Suppose that $\varphi$ is strategy-proof and voter-sovereign. Let us use the curve characterization.

\[\varphi(\vv{r}) := \mbox{sup } \{y | \alpha_{\theta(\vv{r},y)} \geq y \}\]

Then for $y = \min\{r_j\}$, we have $\theta(\vv{r},y)=\vv{\mu}^+$ and $\alpha_{\theta(\vv{r},y)} = \mu^+ \geq y$. Therefore $\varphi(\vv{r}) \geq \min\{r_j\}$.

Similarly, for $y = \max\{r_j\} + \epsilon$, where $\epsilon>0$, we have $\theta(\vv{r},y)=\vv{\mu}^-$ and $\alpha_{\theta(\vv{r},y)} = \mu^- < y$. Therefore $\varphi(\vv{r}) \leq \max\{r_j\}$.

We have shown that $min\{r_j\} = r_k \leq \varphi(\vv{r}) \leq r_l = max\{r_j\}$. 

As such for any voting profile $\vv{s}$ if $\varphi(\vv{s}) < \varphi(\vv{r})$ (resp. $\varphi(\vv{s}) > \varphi(\vv{r})$) then voter $l$ (resp. $k$) has a worse outcome in $\vv{s}$ than in $\vv{r}$ according to peak $r_l$ (resp. $r_k$). It follows that no voting profile can obtain a result that is strictly better for at least one voter without being worse for another.

\end{proof}

\subsection{Strict Responsiveness and Ordinality}

\begin{axiom}[Strict Responsiveness]
A voting rule $\varphi: \Lambda^{N} \rightarrow \Lambda$ is strictly responsive if for any $\vv{r}$ and $\vv{s}$ such that for all $j$, $r_j < s_j$ we have $\varphi(\vv{r}) < \varphi(\vv{s})$.
\end{axiom}

Strict responsiveness is sometimes called strict monotonicity.

\begin{axiom}[Ordinality]
A voting rule $\varphi:\Lambda^{N} \rightarrow \Lambda$ is ordinal if for all strictly responsive and surjective functions $\pi :\Lambda \rightarrow \Lambda$ we have:

\[\varphi(\pi(r_1),\dots,\pi(r_n)) = \pi \circ \varphi(\vv{r}).\]
\end{axiom}

\begin{remark}
Ordinality says nothing when $\Lambda$ is finite. In this case, the identity is the only strictly responsive and surjective function from $\Lambda$ to itself.
\end{remark}

\begin{proposition}\label{prop ordinality}
For a strategy-proof voting rule $\varphi: \Lambda^{N} \rightarrow \Lambda$ the following are equivalent:\\
(1) The phantom function $\alpha$ verifies $\alpha(\Gamma) = \{\mu^-,\mu^+\}$\\
(2) $\varphi$ is strictly responsive.\\
And when moreover $\Lambda$ is rich,\footnote{$\Lambda$ is rich if for any $\alpha < \beta$ in $\Lambda$ there exists a $\gamma\in \Lambda$ such that $\alpha<\gamma < \beta$.} (1) and (2) are equivalent to:\\
(3) $\varphi$ is ordinal and not constant.
\end{proposition}

\begin{proof}
See Appendix \ref{Proof prop ordinality}.
\end{proof}

In the anonymous case, using Moulin's median representation in the proof above, we deduce that the ordinal / strictly monotonic strategy-proof voting rules are the \emph{order functions}, where for $k=1,...,n$, the $k$th-order function is the SP-rule which associates to any input $\vv{r}=(r_1,...,r_n)\in \Lambda^n$ the $k$th-highest value $r_{(k)}$, where $r_{(1)}\geq ... \geq r_{(k)}\geq...\geq r_{(n)}$  (See \cite{BL}, Chapters 10 and 11). Proposition \ref{prop ordinality} can be viewed as a conceptional foundation of the $k$th-order function via an ordinality principle.

\subsection{Dummy Voter}

\begin{axiom}[Dummy Voter]
A voter $i$ is said to be dummy if for all $\vv{r}$ and for all $\vv{s}$ that only differs from $\vv{r}$ in dimension $i$ we have, $\varphi(\vv{r}) = \varphi(\vv{s})$.
\end{axiom}

\begin{proposition}
A voting rule $\varphi$ is strategy-proof and voter $i$ is dummy iff for all $X$ and $Y$ that only differ in $i$, $\alpha_X=\alpha_Y$
\end{proposition}

\begin{proof}
$\Rightarrow:$ Suppose that voter $i$ is dummy then for $X,Y \in \Gamma$ that only differ in $i$:

\[\alpha_X = \varphi(X) = \varphi(Y) = \alpha_Y. \]

$\Leftarrow:$ Suppose that for all $X$ and $Y$ that only differ in dimension $i$, we have $\alpha_X = \alpha_Y$:

Let $\vv{r}$ differ from $\vv{s}$ only in dimension $i$. For all $y$, $\theta(\vv{r},y)$ and $\theta(\vv{s},y)$ differ at most in dimension $i$ therefore: $\alpha_{\theta(\vv{r},y)} = \alpha_{\theta(\vv{s},y)}$. Using the curve representation, it follows that:
\[\varphi(\vv{r}) = \sup \left\{ y \in \Lambda | \alpha_{\theta(\vv{r},y)} \geq y \right\} =  \sup \left\{ y \in \Lambda | \alpha_{\theta(\vv{s},y)} \geq y \right\} = \varphi(\vv{s}).\]
\end{proof}

\subsection{Anonymity}

As seen above anonymity states that all voters must be treated equally. 

\begin{proposition}\label{prop anonymity}
A strategy-proof voting rule $\varphi : \Lambda^{N} \rightarrow \Lambda$ is anonymous iff its phantom function $\alpha$ is anonymous.
\end{proposition}

\begin{proof}
$\Rightarrow:$ It is immediate from (\ref{def_phantom}) that if $\varphi$ is anonymous then so is $\alpha.$

$\Leftarrow:$ For any permutation $\sigma$, let $\vv{s}=(s_1,\dots,s_n) = (r_{\sigma(1)},\dots,r_{\sigma(n)})$ be that permutation of the peaks.

For all $k$ we have that $X_k(\vv{r})$ and $X_k(\vv{s})$ both have $k$ values $\mu^+$ and $n-k$ values $\mu^-$. Therefore since $\alpha$ is anonymous, we have $\alpha(X_k(\vv{r})) = \alpha(X_k(\vv{s}))$. Thus, using the median representation:

\begin{align*}
  \varphi(\vv{s}) &= med(r_{\sigma(1)},\dots,r_{\sigma(n)},\alpha_{X_0(\vv{s})},\alpha_{X_1(\vv{s})},\dots,\alpha_{X_n(\vv{s})}) \\
  &= med(r_1,\dots,r_n,\alpha_{X_0(\vv{r})},\alpha_{X_1(\vv{r})},\dots,\alpha_{X_n(\vv{r})}) \\
  &= \varphi(\vv{r}).
\end{align*}
\end{proof}

\begin{theorem}[Grading Curve Representation. Anonymous Case]\label{lemma-Curve}
A strategy-proof voting rule $\varphi : \Lambda^{N} \rightarrow \Lambda$ is anonymous iff there exists an increasing function $g^n : [0,1] \rightarrow \Lambda$ such that: 
\[\forall X \in \Gamma; \alpha(X) = g^n\left(\frac{\#\mu^+(X) }{n}\right).\]
Where $n$ is the cardinality of $N$.
\end{theorem}

This immediately implies Theorem \ref{functional anymous characterization} above, without the need of assuming $\Lambda$ a compact interval.

\begin{proof}
$\Rightarrow:$ By Proposition \ref{prop anonymity}, $\alpha$ is anonymous, so for each $i$ there is $\alpha_i$ with $\alpha_X = \alpha_i$ for all $X\in \Gamma$ with $\#\mu^+(X)=i$. Therefore we can choose $g^n$ increasing with $g^n(\frac{i}{n}) = \alpha_i$.

$\Leftarrow:$ Suppose $\forall X, \alpha_X = g^n\left(\frac{\#\mu^+(X)}{n}\right)$. It follows that if $X'$ is a permutation of $X$, then $\#\mu^+(X) = \#\mu^+(X')$ and $\alpha_X = \alpha_{X'}$.
\end{proof}

One can avoid the dependency of the grading curve $g^n$ on the number of voters $n$ using a variable electorate approach and imposing consistency conditions (\`a la Smith \cite{Smith} and Young \cite{young}). This is the objective of the next section.

\section{Additional Properties: Variable Electorate}

We wish to consider situations where casting a vote or not is a choice. As such we need to make a distinction between the electorate $\mathcal{E}$ (that is potentially infinite) and the set of voters $N \subseteq \mathcal{E}$ (which will be assumed finite). Henceforth, we define a \emph{voter} as someone who chooses to cast a ballot and an \emph{elector} as someone who can cast a ballot. Similarly a \emph{ballot} is cast by a voter while a \emph{vote} is the response of an elector. A vote that is not a ballot is represented by the symbol $\emptyset$ (interpreted as abstention). As such we represent the set of votes as an element of $\Lambda^* = \{ \vv{r} \in (\Lambda \cup \emptyset)^\mathcal{E} : \# r_i \neq \emptyset < + \infty\}$. 

Since we seek for strategy-proof methods, it may also be of interest to ensure that no elector can benefit from not casting a ballot (the absence of the no-show paradox or participation \cite{Moulin-no-show}). Another desirable property is consistency. It states that when we obtain the same outcome for the voting profiles of two disjoint group of voters (with fixed ballots) then that outcome is also the outcome of the union of their voting profiles. First, we need to extend our definitions and concepts to the variable electorate context.

\begin{definition}[Voting function]
A voting function $\varphi^* : \Lambda^* \rightarrow \Lambda$ is a function such that for any finite set of electors $N$ there is a voting rule $\varphi_N: \Lambda^N \rightarrow \Lambda$ such that $\varphi_N$ is the restriction of $\varphi^*$ to the set of voters $N$.
\end{definition}

Intuitively once our set of voters is fixed then we are considering a voting rule and we can use our previous results. Furthermore in the general case the voting rules are independent even if they use very similar (but not identical) set of voters. As such we start by determining what our set of voters is and then we use the corresponding voting rule. This is coherent with the non-anonymous setting where the result of the election strongly depends on who casts a ballot.

\begin{definition}[Redefining concepts for variable electorate]
A voting function $\varphi^*$ verifies one of the previously mentioned properties (strategy-proofness, continuity, voter sovereignty, dummy voter, strict responsiveness, anonymity) if for all sets of voters $N$ the restriction of $\varphi^*$ to $N$ verifies the property.
\end{definition}

As with voting rules, we would like to define a phantom function that completely characterizes the strategy-proof voting functions. We will then be able to characterize participation (e.g. no no-show paradox) and consistency by using the phantom functions.

Let $\Gamma^* := \{\mu^-,\mu^+,\emptyset\}^\mathcal{E}$ be the set of voting profiles where voters have extreme positions or can abstain. (Recall that the set of voters is always finite). As before we wish for a phantom $\alpha : \Gamma^* \rightarrow \Lambda$ such that:

\[ \forall X \in \Gamma^*, \alpha_X =: \varphi^*(X)\]

The bijection over each set of voters gives us the bijection between a strategy-proof voting function and its associated phantom function. The definition of a phantom function therefore corresponds to all functions that are phantom functions for each voting rule corresponding to a restriction of the voting function to a fixed set of voters.

\begin{definition}[Phantom functions extended to variable electorate]
A function $\alpha : \Gamma^* \rightarrow \overline{\Lambda}$ is a phantom function if $\alpha$ verifies for any $X$ and $Y$ that differ only in dimension $i$ with $X_i = \mu^-$ and $Y_i = \mu^+$ we have $\alpha_X \leq \alpha_Y$.
\end{definition}

\begin{definition}[The $\theta$ function extended to variable electorate]
\begin{align*}
\theta : & (\mathbf{R}\cup \{ \emptyset \})^\mathcal{E} \times \mathbf{R} \rightarrow \{\mu^-,\mu^+,\emptyset\}^\mathcal{E} \\
\theta : & \vv{r},x \rightarrow X=\theta(\vv{r},x)
\end{align*}

Such that $\forall i;~X_i = \mu^- \Leftrightarrow r_i < x$ and $\forall i;~X_i = \mu^+ \Leftrightarrow r_i \geq x$.
It follows that $X_i = \emptyset$ means that the elector $i$ did not vote ($r_i = \emptyset$).
\end{definition}
The intuition behind the $\theta$ function remains the same. We divide the voters into two groups, those whose ballot is greater or equal to $x$ and those whose ballot is lower than $x$.

\begin{remark}
All previous results remain true with the new definitions. In particular the characterizations of strategy-proof voting functions are the same as those of strategy-proof voting rules.
\end{remark}

\subsection{Participation (or No No-Show Paradox)}

\begin{axiom}[Participation]
A voting function $\varphi^* : \Lambda^* \rightarrow \Lambda$ is said to verify participation if for all $\vv{r}$ where $r_i \neq \emptyset$ and $\vv{s}$ that only differs from $\vv{r}$ in dimension $i$ with $s_i = \emptyset$ we have:

\[\varphi^*(\vv{s}) \geq \varphi^*(\vv{r}) \geq r_i\]

or 
\[\varphi^*(\vv{s}) \leq \varphi^*(\vv{r}) \leq r_i.\]

\end{axiom}

Participation is a natural extension of strategy-proofness. ``Strategy-proofness + participation'' is equivalent to no matter what the elector does, no strategy gives a strictly better outcome than an honest ballot.

\begin{theorem}[Participation and Phantom Functions] \label{theo participation}
A strategy-proof voting rule $\varphi^* : \Lambda^* \rightarrow \Lambda$ verifies participation iff with the order $\mu^- < \emptyset < \mu^+$, its associated phantom function $\alpha$ is weakly increasing.
\end{theorem}

\begin{proof}
See Appendix \ref{Proof of theo participation}
\end{proof}

\begin{remark}
Strategy-proofness is obtained by requiring a phantom function to be increasing. Hence, there is no surprise that ``strategy-proofness + participation'' implies that the phantom function is increasing for the order $\mu^- <\emptyset< \mu^+$.
\end{remark}

It is of no surprise that in the voting by issues approach, participation means monotonicity with respect to the addition of a new member.

\begin{proposition}[Participation and Winning Coalitions]\label{prop Participation and Winning Coalitions}
In a vote by issue, a strategy-proof voting function verifies participation iff when a elector $i$ decides to become a voter with ballot $x$ then for any property $H$ containing $x$, if $W_H$ was a winning coalition of $H$ for the initial set of voters then $W_H \cup \{i\}$ is a winning coalition for the new set of voters.
\end{proposition}

\begin{proof}
See Appendix \ref{Proof of prop Participation and Winning Coalitions}.
\end{proof}

\subsection{Consistency}

In order to define consistency, we introduce the function $\sqcup : \Lambda^* \times \Lambda^* \rightarrow \Lambda^*$ that takes the ballots of two disjoint sets of voters and returns the union of the ballots. 

\begin{axiom}[Consistency]
A voting function $\varphi^*$ is said to be {consistent} if for any two disjoint sets of voters $R$ and $S$, when $\vv{r}$ represents the ballots of $R$ and $\vv{s}$ represents the ballots of $S$, we have:

\[\varphi^*(\vv{r}) = \varphi^*(\vv{s}) \Rightarrow \varphi^*(\vv{r}) = \varphi^*(\vv{r} \sqcup \vv{s}).\]
\end{axiom}

\begin{theorem} [Consistent SP Voting Rules]
A strategy-proof voting function $\varphi^* : \Lambda^* \rightarrow \Lambda$ verifies consistency iff for all $X, Y \in \Gamma^*$ with disjoint sets of voters and $\alpha_X \leq \alpha_Y$ we have: 
\[\alpha_X \leq \alpha_{X\sqcup Y}\leq \alpha_Y.\]
\end{theorem}

\begin{proof}
$\Rightarrow:$ By \emph{reductio ad absurdum}. Let us suppose $\alpha_{X \sqcup Y} < \alpha_{X} \leq \alpha_Y$. Let us define $\vv{s}$ as:
\begin{itemize}
    \item $Y_j = \mu^- \Rightarrow s_j = \mu^-$;
    \item $Y_j = \mu^+ \Rightarrow s_j = \alpha_X$.
\end{itemize}
Therefore  $\varphi^*(X) = \alpha_X$, $\varphi^*(\vv{s}) = \alpha_X$ and $\varphi^*(\vv{X \sqcup s}) = \alpha_{X \sqcup Y}$. This contradicts consistency.

A similar proof shows that we cannot have $\alpha_X \leq \alpha_Y < \alpha_{X \sqcup Y}$.

$\Leftarrow:$ Suppose that for all $X, Y \in \Gamma^*$ that correspond to two disjoint sets of voters $N_1$ and $N_2$ such that $\alpha_X \leq \alpha_Y$ we have $\alpha_X \leq \alpha_{X \sqcup Y}\leq \alpha_Y.$

If $\varphi^*(\vv{r}) = \varphi^*(\vv{s})=a$ then:

\begin{itemize}
    \item If $a=r_i=s_k$: then for all $\epsilon > 0$ 
    \[\alpha_{\theta(\vv{r} \sqcup \vv{s},r_i+\epsilon)} \leq max(\alpha_{\theta(\vv{r},r_i+\epsilon)},\alpha_{\theta(\vv{s},r_i+\epsilon)}) \leq r_i\]
    and 
    \[r_i \leq  min(\alpha_{\theta(\vv{r},r_i)},\alpha_{\theta(\vv{s},r_i)}) \leq \alpha_{\theta(\vv{r} \sqcup \vv{s},r_i)}.\]
    For $\epsilon$ small enough $\mu^+(\theta(\vv{r}\sqcup \vv{s} ,r_i+\epsilon))=\{j \in N_1| r_j > r_i \} \cup \{j \in N_2 | s_j > r_i\}$. Therefore $\varphi(\vv{r} \sqcup \vv{s}) = a$.  Consistency is verified.
    
    \item If $a=r_i=\alpha_{X}$ where $X =\vv{s}$, then for all $\epsilon>0$:
    \[\alpha_{\theta(\vv{r},r_i+\epsilon)} \leq \alpha_{\theta(\vv{r} \sqcup X,r_i+\epsilon)} \leq \alpha_{X}=r_i\]
    and 
    \[r_i =\alpha_X \leq \alpha_{\theta(\vv{r} \sqcup X,r_i)} \leq \alpha_{\theta(\vv{r},r_i)}.\]
    For $\epsilon$ small enough $\mu^+(\theta(\vv{r}\sqcup X ,r_i+\epsilon))=\{j \in N_1| r_j > r_i \} \cup \{j \in N_2 | s_j > r_i\}$.
    Therefore $\varphi(\vv{r} \sqcup \vv{s}) = a$. Consistency is verified.
    \item If $a=\alpha_{X}=\alpha_{Y}$ where the voters of $X=\vv{r}$ and the voters of $Y=\vv{s}$: 
    \[\alpha_{X \sqcup Y} = \alpha_X = \alpha_Y.\]
    Therefore  $\varphi(\vv{r} \sqcup \vv{s}) = a$. Consistency is verified.
\end{itemize}

\end{proof}

It is important to notice that the set of wining coalitions in a voting by issue context is defined for each fixed set of voters, but there is no consistency imposed a priori when the set of voters changes. When consistency is imposed, in addition to SP, some monotonicity over the set of winning coalitions is obtained. 

\begin{proposition}[Consistent Voting Coalitions]
In a vote by issues, a strategy-proof voting function verifies consistency iff for any $a\in \Lambda$ if $X$ and $ Y$ $ \in \Gamma^*$ have disjoint sets of voters:
\begin{itemize}
    \item When $\mu^+(X)$ and $\mu^+(Y)$ are winning coalitions for $H = \{y \geq a \}$ then $\mu^+(X) \cup \mu^+(Y)$ is a winning coalition for $H$.
    \item When $\mu^-(X)$ and $\mu^-(Y)$ are winning coalitions for $H = \{y \leq a \}$ then $\mu^-(X) \cup \mu^-(Y)$ is a winning coalition for $H$.
\end{itemize}
\end{proposition}

\begin{proof}
$\Rightarrow:$ Suppose that our strategy-proof voting verifies $\alpha_X \leq \alpha_{X \sqcup Y} \leq \alpha_Y$ for all $X$ and $Y$ that correspond to disjoint sets of voters.

A simple inequality consideration for any $a$ gives the result.

$\Leftarrow:$ Suppose $\alpha_X \leq \alpha_Y$. $\mu^+(X)$ and $\mu^+(Y)$ are winning coalitions for $\{ y \geq \alpha_X\}$ therefore $\alpha_{X \sqcup Y} \geq \alpha_X$. Conversely $\mu^-(X)$ and $\mu^-(Y)$ are winning coalitions for $\{ y \leq \alpha_X\}$ therefore $\alpha_{X \sqcup Y} \leq \alpha_Y$.
Therefore we verify participation.

\end{proof}

\begin{proposition} A strategy-proof voting function  $\varphi^* : \Lambda^* \rightarrow \Lambda$ that verifies voter sovereignty and consistency also verifies participation.
\end{proposition}

\begin{proof}
Let $X \in \Gamma^*$ and $Y \in \Gamma^*$ be the voting profile where only $i$ is a voter and where he votes respectively $\mu^-$ and $\mu^+$.
By voter sovereignty, $\alpha_X =\mu^-$ and $\alpha_Y =\mu^+$. By consistency, for any $Z \in \Gamma^*$ where $i$ is not a voter: 

\[ \alpha_X \leq \alpha_{X \sqcup Z} \leq \alpha_Z \leq \alpha_{Y \sqcup Z} \leq \alpha_Y\]
\end{proof}

\subsection{Properties Summary Table}\label{additional properties}

\begin{center}
\begin{tabular}{|l|l|l|l|} 
   \hline
    Properties & Phantom function & Winning coalitions \\ 
    \hline
    Voter sovereignty & $\alpha(\vv{\mu}^-) = \mu^-$ & $N \in \mathcal{W}_{\{y \geq \mu^+\}}$\\
     & $\alpha(\vv{\mu}^+) = \mu^+$ & $N \in \mathcal{W}_{\{y \leq \mu^-\}}$ \\ 
    \hline 
       Strict Responsiveness & $\alpha(\Gamma) = \{\mu^-,\mu^+\}$ & $\forall \vv{r}, \exists r_i, $ \\
       Ordinality & & $\{r_j \leq r_i\} \in \mathcal{W}_{\{y \leq r_j\}}$ \\
        & & $\{r_j \geq r_i\} \in \mathcal{W}_{\{y \geq r_j\}}$ \\
    \hline
       Dummy voter $i$ & $(\forall j \neq i, X_j = Y_j)$ & $\forall H \in \mathcal{H}, W \in \mathcal{W}_H$ \\
        & $\Rightarrow$ & $\Rightarrow$ \\
        & $\alpha(X) = \alpha(Y)$ & $ W - \{i\} \in \mathcal{W}_H $\\
        \hline
        Anonymity & $\forall X ,Y: \sum_i X_i = \sum_i Y_i$ & $\forall H \in \mathcal{H}, \exists p,$ \\
        & $\Rightarrow$ & $\mathcal{W}_H = \{ W |\#W \geq p\}$ \\
        & $\alpha(X) = \alpha(Y)$ &  \\
        \hline
        Participation & $\alpha$ is monotonous & $ W \in \mathcal{W}_H$\\
        & & $\Rightarrow  W \cup \{i\} \in \mathcal{W}_H$\\
        \hline
        Consistency & $\alpha_X \leq \alpha_Y \wedge \forall j, \emptyset \in \{X_j,Y_j\} \Rightarrow$ & $W_1, W_2 \in \mathcal{W}_H$\\
        & $\alpha_X \leq \alpha_{X \sqcup Y} \leq \alpha_Y$ & $W_1 \cup W_2 \in \mathcal{W}_H$\\
        \hline

\end{tabular}
\end{center}

\subsection{Combining Consistency and Anonymity}

In this section we show that consistency in the anonymous case is equivalent to removing the dependency of the grading curves on the number of voters $n$.

\begin{axiom}[Homogeneity]
A voting profile is homogeneous if for any two profiles $\vv{r}$ and $\vv{s}$ such that there exists $k\geq 1$ that verifies:
\[\forall x\in\Lambda, \#\{j |x =r_j \in \vv{r}\} = k\#\{ j |x=r_j \in \vv{s}\}\]
we have
\[ \varphi(\vv{s})= \varphi(\vv{r}) .\]
\end{axiom}

Hence, a voting function is homogeneous if for any $k \geq 1$ when each ballot is replaced by $k$ copies of that ballot the result of the function does not change.

\begin{proposition}
A strategy-proof voting function is consistent and anonymous iff it is homogeneous.
\end{proposition}

\begin{proof}
$\Rightarrow:$ Immediate due to the fact that $\alpha_X$ only depends on the fraction $\frac{\# \mu^+(X)}{\# \mu^-(X)+\# \mu^+(X)}$.

$\Leftarrow:$ Suppose that we have an homogenous strategy-proof voting function. By definition this implies anonymity. For any $X$ and $Y$, we can duplicate in order to have $X'$, $Y'$ and $(X \sqcup Y)'$ with the same number of voters in each and $\alpha_Z = \alpha_{Z'}$ for $Z \in \{X,Y,X \sqcup Y\}$. By barycentric considerations:
\[\mu^+(X') \leq \mu^+((X \sqcup Y)') \leq \mu^+(Y') \]

Therefore by definition of a phantom function $\alpha_X \leq \alpha_{X \sqcup Y} \leq \alpha_Y$.
\end{proof}

\begin{theorem}[Consistent Grading Curves] \label{Curve and Consistency}
A SP voting function $\varphi^*=(\varphi^n) : \Lambda^* \rightarrow \Lambda$ is anonymous and consistent iff there is an increasing function $g :[0,1] \rightarrow \Lambda$ (electorate size independent) and a constant $x \in \Lambda$ such that the phantom function $\alpha :\Gamma^*$ is defined as:

\[\alpha_X :=\left\{ \begin{array}{lll}
		g\left(\dfrac{\# \mu^+(X)}{\# N}\right) & \mbox{ if } & \# N \neq 0\\
		x & \mbox{ if } & \# N = 0
    \end{array}
    \right.\]
    
Furthermore the voting function verifies participation iff $x \in g([0,1])$.
\end{theorem}

This is an elegant new result. It says that consistency and anonymity are equivalent to the grading curve being independent of the electorate size. 

\begin{proof}
See Appendix \ref{Proof of Theorem Curve and Consistency}.
\end{proof}

\begin{axiom}[Continuity with Respect to New Members]\cite{Smith,young}.
A voting function $\varphi^*$ is said to be continuous with respect to new members if:
\[ \forall \vv{r},\vv{s},\mbox{lim}_{n \rightarrow +\infty} \varphi^*(\overbrace{\vv{r} \sqcup \dots \sqcup \vv{r}}^n \sqcup \vv{s}) = \varphi(\vv{r})\]
\end{axiom}

Not surprisingly, but elegantly, continuity with respect to new members is equivalent to continuity of the grading curve:

\begin{theorem}[Continuous Grading Curves]\label{Theo Continuous Grading Curves}
A strategy-proof, homogeneous ($=$ consistent and anonymous) voting function $\varphi^* : \Lambda^* \rightarrow \Lambda$ is continuous with respect to new members iff its grading curve $g$ is continuous.
\end{theorem}

\subsection{Proportionality}

In this subsection we assume that $\Lambda=[0,1]$.

\begin{definition}[Linear=Uniform Median]
The strategy-proof voting function $\varphi^* : [0,1]^* \rightarrow [0,1]$ defined for any $n = \# N$ and $X \in \{0,1\}^{N}$, by $\alpha(X) = \sum_i \frac{X_i}{n}$ is called the linear median. It corresponds to the grading curve $g(x)=x$.
\end{definition}

The linear median was first proposed and studied in the unpublished PhD dissertation \cite{Jennings} of our co-author A. Jennings. It was rediscovered independently by Caragiannis et al. \cite{ICML2016} for its nice statistical properties under the name `uniform median.' These authors use a different representation based on the Moulin phantom characterization. Namely, if there are $n$ voters and $\Lambda=[0,1]$ then the linear median can be computed via the Moulin median formula: \[\varphi(\vv{r})=med(r_1,...,r_n,\alpha_0,...,\alpha_n),\]
where $\alpha_k=\frac{k}{n}, \forall k=0,...,n$, that is, the Moulin $n+1$ phantom voters are uniformly distributed on the interval $[0,1]$. It is not evident from the median representation why this `uniform median' function satisfies participation, consistency, or continuity. But these properties are immediate consequences of the linear grading curve $g(x)=x$ representation since it is continuous and independent of the size of the electorate. We thus obtain:

\begin{proposition}
The linear median satisfies anonymity, continuity, consistency, sovereignty, participation and continuity with respect to new members. On the other hand, it is neither ordinal nor strictly responsive.
\end{proposition}

The proof is trivial from previous subsections as $g$ is independent on the electorate size, is continuous, is onto, etc. 

\begin{axiom}[Proportionality]
A voting function $\varphi^* : [0,1]^* \rightarrow [0,1]$ is proportional if 
\[\forall X \in \{0,1\}^{N}, \varphi^*(X) = \frac{\sum_i X_i}{\# N}.\]
\end{axiom}

\begin{theorem}
A voting function $\varphi^* : [0,1]^* \rightarrow [0,1]$ is strategy-proof and proportional iff it is the linear median.
\end{theorem}

This a direct consequence of the fact that a SP function is completely determined by its phantom function as proved in Theorem \ref{ICarac}.  

\begin{remark}
Freeman et al.~\cite{Freeman} showed that the uniform median is the unique proportional, anonymous and continuous strategy-proof method (even in a more general multi-dimensional setting). From our characterization it follows that anonymity is not necessary in the one-dimensional case (nor continuity, because we consider a variable electorate, while they considered a fixed electorate).
\end{remark}

Here are the various characterizations of the linear median, depending on the representation, where $n$ is the electorate size. 

{\small
\begin{center}\label{table charac linear}
    \begin{tabular}{|l|l|}
    \hline
         Characterization & Formula\\
         \hline
        Phantom functions & $\varphi(\vv{r}) := \left\{ \begin{array}{lll}
		\dfrac{k}{n} & \mbox{ if } & \#\left\{ j | \dfrac{k}{n} \leq r_j \right\} = k\\
		& & \\
		r_i & \mbox{ if } &  \dfrac{k-1}{n} \leq r_i \leq \dfrac{k}{N} \mbox{ where } \#\{ j | r_i < r_j \} = k
    \end{array}
    \right. $  \\
        \hline
        &  \\
        MaxMin & $\varphi(\vv{r}) =\sup_{S \subset N}  \min{\left(\dfrac{\#S}{n}, \inf_{i\in S} \{ r_i\}\right)}$\\
        & \\
        \hline
        Phantom characterization & $\varphi(\vv{r}) := med\left(r_1,\dots,r_n,\dfrac{1}{n},\dots,\dfrac{n-1}{n}\right)$\\
       & \\
        \hline 
        Functional characterization & $\varphi(\vv{r}) := \mbox{sup } \left\{y | \dfrac{\#\{r_j \geq y \}}{n} \geq y \right\}$ \\
        & \\
        \hline
        & \\
        Voting by issues & $\varphi(\vv{r}) := a$  :  $\#\{r_i \leq a\} \geq \lceil (1-a)n \rceil \wedge \{r_i \geq a\} \geq \lfloor an \rfloor$ \\ 
        &   \\
        \hline
    \end{tabular}
\end{center}

}

\section{Maximizing Social Welfare}

In economics, social welfare is often taken to be the sum of individual utilities. This section deals with the maximization of social welfare under the strategy-proofness constraint. We will measure the individual utilities by the $L_q$-distance to the peaks and our objective is to compute the SP socially optimal mechanism for each $q\in [1,+\infty]$.

\begin{definition}[Ex-Post Social Welfare]
The ex-post social welfare for a given voting rule $\varphi : \Lambda^n \rightarrow \Lambda$ is defined to be:
$$SW(\varphi,\vv{r}):= - \sum_i \| \varphi(\vv{r}) - r_i \|_q$$
for a given norm $L_q$.
\end{definition}

On all the rest of the section, we will assume that $\Lambda = [m,M]$. 

\subsection{Maximizing Ex-Post Welfare Without the SP Constraint}

Let us start by computing the voting rule that maximizes the ex-post social welfare, without imposing the SP constraint. For the $L_1$-norm, the solution is happily an SP rule, but this is not the case for the other norms.

\begin{theorem}[The Social Welfare Maximizer under the $L_1$-Norm]
A voting rule maximizes Social Welfare for the $L_1$-norm iff there exists a fixed value $\alpha \in \Lambda= [m,M]$ such that :
\begin{equation*}
\label{L_1 welfare}
\forall \vv{r}; \varphi(\vv{r}) := \left\{ \begin{array}{ll}
		med(r_1,\dots,r_n) & \mbox{ if n is odd} \\
        med(r_1,\dots,r_n,\alpha) & \mbox{ if n is even } 
    \end{array}
    \right.
\end{equation*}
\end{theorem}

\begin{proof}
This is a well known result, see for instance \cite{BL}, Section 12.4.
\end{proof}

This estimator is already an anonymous and voter sovereign SP function. It can easily be extended into a consistent voting function (use the same $\alpha$ for all even numbered electorates).  On the other hand, it is trivial to establish that the unique voting rule that always maximizes the $L_2$ ex-post social welfare is the arithmetic mean : $\phi(r_1,...,r_n)=\frac{1}{n}\sum_i^n r_i$ -- which is clearly not a SP voting rule. This extends to all $L_q$-norms for $q>1$:

\begin{theorem}
For any $q>1$,and for any $\# N \geq 2$ there is a unique anonymous voting rule that always minimizes the $L_q$-distance between the inputs and the output. It is not SP.
\end{theorem}

\begin{proof}
See the Appendix \ref{Proof Ex-Post Welfare} where the statement is proved in the more general case where voters have weights. For $q=\infty$, the optimal voting rule is the arithmetic mean of the min and max of the inputs.  This voting rule is easily manipulable.

\end{proof}

\subsection{Maximizing the Ex-Ante Welfare under the SP Constraint}

It is not difficult to see that no fixed SP voting rule is ex-post optimal for every realization $\vv{r}$. Hence we will (as it is done usually) optimize ex-ante (given some prior distribution) or adversarially (under the worst possible scenario). 

To calculate the ex-ante optimal SP voting functions, we need a probability distribution $p$ for the inputs and we should integrate over $[m,M]^n$, weighted by that probability distribution.  Moreover, it is possible to choose a different norm for the integration than the norm for measuring utilities. Hence, in the general case, for fixed $q,r \geq 1$, we seek to find the SP voting rule $f$ which, minimizes

\[\left(\int_m^M \cdots\int_m^M \left(\sum_{i=1}^n |x_i-f(\vv{x})|^q\right)^{\frac{r}{q}} \prod_{i=1}^n p(x_i) dx_n\ldots dx_1\right)^{\frac{1}{r}}\]

We do not solve this in the general case, but only consider the special case where $r=q$ (and in last section, the case where $r=+\infty$).  

Let $p$ be a probability distribution on $\mathbf{R}$ with support $\Lambda = [m,M]$. For $q\geq 1$, we define $G_p^q$ on $(m,M]$ as:
\[G_p^q(x) = \left( 1 + \frac{\frac{\int_x^M (t-x)^{q-1} p(t) dt}{\int_x^M p(t) dt}}{\frac{\int_m^x (x-t)^{q-1} p(t) dt}{\int_m^x p(t) dt}} \right)^{-1}\]

It is clear that $G_p^q(M) = 1$ and $\lim_{x\rightarrow m^+} G_p^q(x) = 0$.  Also, $G_p^q$ is continuous by construction, so we extend $G_p^q$ continuously to $G_p^q(m)=0$.  If $G_p^q$ is strictly increasing, then it is invertible and we define the inverse $g_p^q : [0,1] \rightarrow [m,M]$.

\begin{theorem}\label{norm-minimize}
Let $p$ be a probability distribution on $\mathbf{R}$ with support $\Lambda = [m,M]$. If $G_p^q$ is strictly increasing, then the grading curve $ g_p^q$ generates the unique  anonymous and voter-sovereign SP voting function which minimizes the $L^q$-norm between inputs and output, integrated with the $L^q$-norm over all inputs (i.i.d. from $p$).
\end{theorem}

\begin{proof}
See the Appendix \ref{Proof Ex-Ante Welfare} where the statement is proved in the more general case where voters may have (non equal) weights.
\end{proof}

For the special case where all weights are equal to 1, $q = 1$ and any distribution $p$, this gives
\[G_p^1(x) = \frac{1}{2}\]
Therefore the grading curve verifies:

\[g_p^1(y) = \left\{ \begin{array}{ll}
		0 & \mbox{ if y $<$ 0.5} \\
        1 & \mbox{ if y $>$ 0.5} \\
        \alpha & \mbox{if y = 0.5}
    \end{array}
    \right.\]
 This is the grading curve for the voting function that maximizes social welfare when considering the $L_1$-norm.

\subsubsection{For $q = 2$:}

\[G_p^2(x) = \frac{x - \frac{\int_m^x t p(t) dt}{\int_m^x p(t) dt}}{\frac{\int_x^M t p(t) dt}{\int_x^M p(t) dt} - \frac{\int_m^x t p(t) dt}{\int_m^x p(t) dt}}\]

We can rearrange the above equation to:

\[x = G_p^2(x)\frac{\int_x^M t p(t) dt}{\int_x^M p(t) dt} + (1 - G_p^2(x))\frac{\int_m^x t p(t) dt}{\int_m^x p(t) dt}\]

To give an intuition, we note that $\frac{\int_m^x t p(t) dt}{\int_m^x p(t) dt}$ is the expected value of a number from $p$ given that it is less than $x$ and $\frac{\int_x^M t p(t) dt}{\int_x^M p(t) dt}$ is the expected value of a number from $p$ greater than $x$.  

Hence, $G_p^2(x)$ gives the number $y \in (0,1)$ such that if $y$ is the fraction of the inputs greater than $x$ then $x$ is the expected value of the arithmetic mean of the inputs.  Thus $G_p^2$ is, in some sense, an attempt to emulate the arithmetic mean with an SP voting function.

\subsubsection{The uniform case:}

If the probability distribution $p$ we use is the uniform distribution $U$ on the interval $[m,M]$.  Then,

\[G_U^q(x) = \left( 1 + \left(\frac{M-x}{x-m}\right)^{q-1}\right)^{-1}\]

and 
\[g_U^q(y) = m + (M-m)\left(1 + \left(\frac{1}{y} - 1\right)^\frac{1}{q-1}\right)^{-1}\]

Which is increasing. Consequently:

\begin{theorem}\label{Welfare Uniform Prior}
 When $p$ is the uniform distribution, $g_U^q$ generates the grading curve of the unique SP anonymous onto rule maximizing the ex-ante $L_q$-welfare.
\end{theorem}

Of particular note is $q=2$ which gives the grading curve $g_U^2 : y \rightarrow (M-m)y+m$ and when $m=0$ and $M=1$, then $g_U^2(y)=y$. Therefore:

\begin{theorem}
When $[m,M]=[0,1]$ and $p$ is the uniform distribution, the linear median is  the unique SP anonymous onto rule maximizing the ex-ante $L_2$-welfare.
\end{theorem} 

\subsection{ Maximizing the Ex-ante Welfare with an Adversarial Prior}

Another interesting set of cases to consider is when the outer norm is minimax ($r=\infty$).  In this case, we seek to find the voter-sovereign, anonymous and strategy-proof voting rule $f$ that minimizes:

\[\max_{\vv{x}\in\mathbf{R}^n} \sum_{i=1}^n |x_i-f(\vv{x})|^q\]

\begin{theorem}
For $q\ge 1$ and a compact interval $[m,M]$, for all $n$,

\[g_U^q(x) = m + (M-m)\left(1 + \left(\frac{1}{x} - 1\right)^\frac{1}{q-1}\right)^{-1}\]
is the unique grading curve that generates the voter-sovereign, anonymous and strategy-proof voting rule which minimizes the maximum possible $L_q$-norm between inputs in $[m,M]^n$ and the output.
\end{theorem}

\begin{proof}
See the Appendix \ref{Proof MinMax Welfare} where the result is proved when voters have weights. \end{proof}

Hence, the social welfare maximizer for the `minmax norm' corresponds to the maximizer with the uniform prior (Theorem \ref{Welfare Uniform Prior}). When $q=2$ and $[m,M]=[0,1]$ the optimal SP rule is the linear (=uniform) median, which is one of the main results in \cite{ICML2016}. Hence, our last theorem is an extension of \cite{ICML2016} for $q\neq 2$ as well as when voters have weights (see Appendix \ref{Proof MinMax Welfare}).

We finish this section (and the paper) with the following open question. For the uniform distribution, since $G_U^q$ is optimal for the $L^\infty L^q$-norm and for the $L^q L^q$-norm then is it optimal for everything in between (for $L^r L^q$ for $r \in [q,\infty)$)?  Perhaps it is even optimal for all $r \in (1,\infty)$.

\section{Conclusion}

We introduced the notions of phantom functions and grading curves and demonstrated their usefulness in (i) unifying a number of existing characterizations of strategy-proof voting rules on the domain of single-peaked preferences, and (ii) obtaining insightful new characterizations. 

As an important example, we have characterized the linear median as the unique strategy-proof voting rule satisfying proportionality or maximizing the ex-ante social welfare under the $L_2$-norm and a uniform ex-ante prior. It has been shown to possess further salient properties such as consistency and participation (because its grading curve, the identity, is size electorate independent). On the other hand, the linear median presupposes a cardinal scale. But adding the natural condition of ordinality characterizes, in the anonymous case, the class of order (statistics) functions which play an important role in the majority judgment -- ordinal -- method of voting (Balinski-Laraki \cite{BL} Chapters 10-13). A particularly appealing order function is the middlemost (Chapter 13 in \cite{BL}). 

Finally, in the appendix it is shown how our model and results extend to multi-dimensional separable input spaces, or to the case where voters have weights.

\bibliographystyle{plain}
\bibliography{SP_in_Single-Peaked}

\newpage
\appendix

\section{Extension to Weighted Voters}

We can also be interested in voting rules where each voter is given a weight. For example the weight of a shareholder vote can be proportional to the number of shares he owns.
We may want to understand in a more general setup (and not only under the proportionality axiom) the effect of attributing weights to electors. This is the objective of this section.

\begin{definition}[Integer Weighted Voters]
The voting function $\varphi : \Lambda^* \rightarrow \Lambda$ is integer weighted with weight sequence $\vv{w}$ if there is a homogeneous voting function $\phi : \Lambda^* \rightarrow \Lambda$ such that:
\[\forall \vv{r}, \varphi(\vv{r}) = \phi(\overbrace{r_1,\dots,r_1}^{w_1 },\dots,\overbrace{r_n,\dots,r_n}^{w_n})\]
where $n$ denotes the last voter.
\end{definition}

\begin{remark}
According to the previous section if $\varphi$ is strategy-proof and weighted with weights $\vv{w}$, there exists a grading curve $g : [0,1] \rightarrow \Lambda$ such that for any $X$ with at least one voter, we have:
\[ \alpha_X = g\left(\frac{\sum_{i \in \mu^+(X)} w_i}{\sum_{i \in N} w_i}\right) \]
\end{remark}

\begin{lemma}
If a strategy-proof voting function $\varphi : \Lambda^* \rightarrow \Lambda$ is integer weighted then it is consistent.
\end{lemma}

\begin{proof}
Suppose that we have $X,Y \in \Gamma^*$ with $\alpha_X \leq \alpha_Y$ such that $X$ and $Y$ are associated to two disjoint sets of voters. Let $\vv{x}, \vv{y}$ be the respective weight vectors.

By barycentric values:
\[\frac{\sum_{i \in \mu^+(X)} x_i}{\sum_{i \in \mu^-(X) \cup \mu^+(X)} x_i} \leq \frac{\sum_{i \in \mu^+(X)} x_i + \sum_{i \in \mu^+(Y)} y_i}{\sum_{i \in \mu^-(X) \cup \mu^+(X)} x_i + \sum_{i \in \mu^-(Y) \cup \mu^+(Y)} y_i} \leq \frac{\sum_{i \in \mu^+(Y)} y_i}{\sum_{i \in \mu^-(Y) \cup \mu^+(Y)} y_i}\]

Therefore $\alpha_X \leq \alpha_{X \sqcup Y} \leq \alpha_Y$.
\end{proof}

It could be interesting to study the impact when voters weights can change.

\begin{definition}[Voting Function with Variable Weights]
A voting function with variable weights $\varphi: \Lambda^* \times \mathbf{N}^\mathcal{E} \rightarrow \Lambda$ is a function such that there is an homogeneous voting rule $\phi$ such that:

\[\forall \vv{w}, \forall \vv{r}, \varphi(\vv{r},\vv{w}) = \phi(\overbrace{r_1,\dots,r_1}^{w_1 },\dots,\overbrace{r_n,\dots,r_n}^{w_n})\]

\end{definition}

\begin{remark}
Our definition ensures that raising the weight of a voter by $1$ is the same as adding a voter of weight $1$ with the same ballot.
\end{remark}

Using homogeneity, one can extend a voting function with integer weights to a voting function with rational weights. The extension to real weights can be obtained using continuity.

\begin{definition}[Real Weighted Voters]
A voting function $\varphi : \Lambda^* \times \mathbf{R}^\mathcal{E} \rightarrow \Lambda$ is a real weighted voting rule with variable weights if there exists a voting rule with variable rational weights $\phi : \Lambda^* \times \mathbf{Q}^\mathcal{E} \rightarrow \Lambda$ such that:
\[ \forall \vv{r}, \lim_{\vv{w} \rightarrow \vv{w'}} \phi(\vv{r},\vv{w}) = \varphi(\vv{r},\vv{w}).\]
\end{definition}

\begin{theorem}[Characterizing Real Weighted Voting Rules]
A strategy-proof voting function with variable weights $\varphi : \Lambda^* \times \mathbf{R}^\mathbf{E} \rightarrow \Lambda$ is well-defined for all real weights iff the grading curve $g$ is continuous in all irrational values.
\end{theorem}

\begin{proof}
Suppose that $g$ is discontinuous in $a\in \mathbf{R} - \mathcal{Q}$ Consider two voters with weights $a$ and $1-a$ where the first votes $M$ and the second $m$. Then $\varphi$ is not well defined since $\lim_{\vv{w} \rightarrow \vv{w'}} \phi(\vv{r},\vv{w})$ does not result when we are nearing $a$ from above or from below. As such $g$ is continuous in all real values. Since $g$ is increasing this ensures that $\varphi$ is well-defined.
\end{proof}

A continuous grading curve can therefore be used to define a voting function for any choice of weights. Consequently:

\begin{theorem}[Grading Curve Representation with Weights]
A voting function $\varphi$ with variable weights is strategy-proof iff there exists a grading curve $g$ such that when there is at least one voter with positive weight:
\[\forall \vv{w} \vv{r}, \varphi(\vv{r},\vv{w}) = \sup\left\{y|g\left(\frac{\sum_{i\in N} w_i 1_{\{ r_i \geq y\}}}{\sum_{i\in N} w_i}\right) \geq y\right\}.\]
\end{theorem}

\begin{axiom}[Proportionality with Weights]
For $\Lambda = [0,1]$, a voting function $\phi^* : \Lambda^* \rightarrow \Lambda$ verifies proportionality for weights $\vv{w}$ if for all $X\in \Gamma^*$
\[\alpha_X = \dfrac{\sum_{i\in N} w_i 1_{\{ r_i \geq y\}}}{\sum_{i\in N} w_i}.\]
\end{axiom}

\begin{theorem}[Linear Median with Weights]
The unique grading curve for the strategy-proof voting function that is proportional with weights is the identity.
\end{theorem}


\section{Extension to the Multi-Dimension Separable $\Lambda$}
We now wish to study the problem of sets of greater dimensions.
Barber\`a, Gul and Stacchetti's \cite{Barbera} model provides us with the results that allow us to do so.

Let $\|.\|$ be a norm over $\mathbf{R}^n$.

Let $\Omega = \Lambda_1 \times \dots \times \Lambda_p$ such that for all $j$, $\Lambda_j \subseteq \mathbf{R}$.

Let $u_i : \Omega \rightarrow \mathbf{R}^+$ be the utility function of a voter $i$. We have that $i$ prefers alternative $X \in \Omega$ to alternative $Y \in \Omega$ iff $u_i(X) \geq u_i(Y)$.

\begin{definition}[Single-peakness]
A utility provides a single-peaked preference on $\Omega$ with peak $A$ if:

\[\forall X,Y \in \Omega; X \in [A,Y] \Rightarrow u_i(X) \geq u_i(Y)\]
\end{definition}
The intuition between our definition of single-peakness is that if $X$ is between my peak $A$ and $Y$ then I weakly prefer $X$ to $Y$ since no matter what criteria I consider as most important, $X$ is at least as good as $Y$.

\begin{axiom}[Strategy-proofness]
For all voters $i$, let $\vv{S}$ that differs from $\vv{R}$ only for voter $i$. Then if $R_i \in \Omega$ is the peak of voter $i$ then $u_i(\varphi(\vv{R})) \geq u_i(\varphi(\vv{S}))$.
\end{axiom}

\begin{lemma} \label{multi ineq}
If $A,X,Y \in \Omega$ are not aligned, then there is a utility function $u$ with peak $A$ such that $u_i(X) > u_i(Y)$.
\end{lemma}

\begin{proof}
Let us consider a basis $e = (e_1,\dots,e_n)$ of $\mathbf{R}^n$ that start with $e_1 = \vv{AX}$, $e_2= \vv{AY}$. Let $u$ be defined as a linear function that verifies $u(e_1) =-2$ and $\forall j \neq 1, u(e_j) = -1$. 

For any $B, C$, if $B \in [AC]$, then $u(B) = \frac{u(C).\|\vv{AB}\|}{\|\vv{AC}\|} \geq u(C)$. Therefore $u$ is single-peaked with peak $A$.

\end{proof}

\begin{lemma} \label{equiv multi}
If $\varphi : \Omega^n \rightarrow \Omega$ is a voting rule, then the two following are equivalent:
\begin{itemize}
    \item $\exists, f_1,\dots,f_n, \forall \vv{R}, \varphi(\vv{R}) = (f_1(R_{1,1},\dots,R_{n,1}),\dots,f_p(R_{1,p},\dots,R_{n,p}))$
    \item For all $j$ and $i$, if $\vv{R}$ and $\vv{S}$ differ only in dimension $i$ and $R_{i,j} = S_{i,j}$ then  $\varphi(\vv{R})_j = \varphi(\vv{S})_j$.
\end{itemize}
\end{lemma}

\begin{proof}
$\Rightarrow:$ Since $\varphi(\vv{R})_j = f_j(R_{1,j},\dots,R_{n,j})$ and $S_{i,j} = R_{i,j}$ for all $i$ we obtain $\varphi(\vv{R})_j = \varphi(\vv{S})_j$

$\Leftarrow:$ Suppose that for any voter $i$, for any $j$ and any set of votes $\vv{R}$ and $\vv{S}$, if $\vv{R}$ and $\vv{S}$ only differ in dimension $i$ and $R_{i,j} = S_{i,j}$ then  $\varphi(\vv{R})_j = \varphi(\vv{S})_j$. Then for any $\vv{T}$ we can by using this equality on one voter at a time, show that if $R_{i,j} = T_{i,j} $  for all $i$ then $\varphi(\vv{R})_j = \varphi(\vv{T})_j$, thus the existence of $f_j$

\end{proof}

\begin{theorem}[Separability]
If $\varphi : \Omega^n \rightarrow \Omega$ is a strategy-proof voting rule then there exists $f_1,\dots,f_p$ where $f_j : \Lambda_j^n \Rightarrow \Lambda_j$ is strategy-proof such that:

$$\forall \vv{R}, \varphi(\vv{R}) = (f_1(R_{1,1},\dots,R_{n,1}),\dots,f_p(R_{1,p},\dots,R_{n,p}))$$ 
\end{theorem}

\begin{proof}
$\Rightarrow:$ According to lemma \ref{equiv multi}. We have that:
\[\forall \vv{R}, \varphi(\vv{R}) = (f_1(R_{1,1},\dots,R_{n,1}),\dots,f_p(R_{1,p},\dots,R_{n,p}))\] is equivalent to, for all $k$ and $i$, if $\vv{R}$ and $\vv{S}$ differ only in dimension $i$ and $R_{i,k} = S_{i,k}$ then  $\varphi(\vv{R})_k = \varphi(\vv{S})_k$.

 Let us take $\vv{R}$ and $\vv{S}$ that differ only in dimension $i$ with $R_{i,k} = S_{i,k}$. Let $\varphi(\vv{R}) = A$ and $\varphi(\vv{S}) =B$.
 
 If $R_i$, $A$ and $B$ are not aligned then according to the lemma \ref{multi ineq} there is a utility function $u$ such that $R_i$ is the peak and $B$ is strictly preferred to $A$. Since we can have $u =u_i$, this contradicts strategy-proofness.

Therefore $R_i$, $S_i$, $A$ and $B$ are aligned. Since $R_{i,k} = S_{i,k}$, the line they are on belongs to the hyperplane $\{Y | Y_k = R_{i,k}\}$. As such $A_k = B_k$.

$\Leftarrow:$ If dimensions are separable and we are strategy-proof dimension by dimension, then we are strategy-proof overall.
\end{proof}

Barber\`a, Gul and Stacchetti's \cite{Barbera} model and results have been extended by Nehring and Puppe \cite{Nehring} \cite{Puppe} to a larger class of combinatorial domains of `generalized single-peaked' preferences. An interesting particular case is the participative budgeting problem in which voters can submit a proposal for how to divide a single divisible resource among several possible alternatives (such as public projects or activities). Under $L_1$-preferences, the social welfare-maximizing mechanism is efficient, anonymous and strategy-proof (see Goel et al.~\cite{Goel}). However, this mechanism fails to be ``proportional'', which leads Freeman, Pennock, Peters and Wortman-Vaughan \cite{Freeman} to propose a new class of moving phantom mechanisms that contains a proportional one. In the particular context of budgeting with only two alternatives, this reduces to the proportional median which they characterize as the unique strategy-proof, anonymous and continuous voting rule satisfying proportionality. As shown above, anonymity and continuity axioms are redundant and that function may be characterized by a linear curve, electorate size independent, implying some of its salient properties such as its consistency, and participation.\footnote{Consistency and participation (or the absence of the no-show paradox) have been proven by  Freeman, Pennock, Peters and Wortman-Vaughan \cite{Freeman}) for the proportional median rules. Our characterization allows understanding the reasons for these.}

The  Freeman, Pennock, Peters and Wortman-Vaughan \cite{Freeman} paper goes much further than the scope of our paper by proposing an extension of the proportional median to budgeting problems with more than two alternatives. However, they do not provide an axiomatization of their proportional method nor do they prove that their moving phantom mechanisms are the unique anonymous and strategy-proof methods. Those questions remain open and we hope our results will help solve them.


\section{Proof of Lemma \ref{continuous extension}}\label{Proof of Lemma continuous extension}

If a voting rule $\varphi : \Lambda^N \rightarrow \Lambda$ is strategy-proof, then it has a unique continuous extension in $\overline{\Lambda}^N \rightarrow \overline{\Lambda}$.

\begin{proof}
We prove the result by induction with respect to the set of players.

It is immediate that we can extend any strategy-proof $\varphi : \Lambda^N \rightarrow \Lambda$ to $\Lambda^N \rightarrow \overline{\Lambda}$.

Suppose that for a given set $S \subset N$ we have shown that we can extend $\varphi : \Lambda^N \rightarrow \Lambda$ to $\overline{\Lambda}^{S} \times \Lambda^{N -S} \rightarrow \overline{\Lambda}$.
We will now show that for any $i \in N - S$ we can extend $\overline{\Lambda}^{S} \times \Lambda^{N -S} \rightarrow \overline{\Lambda}$ to $\overline{\Lambda}^{S \cup \{i\}} \times \Lambda^{N - S - \{i\}} \rightarrow \overline{\Lambda}$.

In the following $\vv{r}$ and $\vv{s}$ have fixed values in all dimensions except for $i$ and only differ in dimension $i$ and $r_i > s_i \geq \mu^-$. We seek to show that $\varphi$ is well defined, by continuity, when $s_i = \mu^-$. The proof for $s_i = \mu^+$ is symmetrical.

\begin{itemize}
    \item  If $r_i < \varphi(\vv{r})$ then by strategy-proofness $\varphi(\vv{r}) \leq \varphi(\vv{s})$. However, when $\mu^- < s_i$, by Lemma \ref{responsiveness} (weak responsiveness), since $s_i < r_i$,  we have $\varphi(\vv{s}) \leq \varphi(\vv{r})$. As such $\varphi(\vv{r}) = \varphi(\vv{s})$. It follows that
\[ \lim_{s_i \rightarrow \mu^-} \varphi(\vv{s}) = \varphi(\vv{r}).\] 
Therefore we can extend by continuity $\varphi(\vv{s})$ to $s_i = \mu^-$.
\item If there exists $x \in \Lambda$ such that $r_i = \varphi(\vv{r})$ for all $r_i < x$, then:
it follows that
\[ \lim_{s_i \rightarrow \mu^-} \varphi(\vv{s}) = \mu^-.\] 
Thus we can extend by continuity $\varphi(\vv{s})$ to $s_i = \mu^-$.
\item If $\varphi(\vv{r}) = \mu^-$, then by Lemma \ref{responsiveness}, when $\mu^- < s_i$, $\mu^- \leq \varphi(\vv{s}) \leq \varphi(\vv{r})= \mu^-$. It follows that
\[ \lim_{s_i \rightarrow \mu^-} \varphi(\vv{s}) = \mu^-.\] 
Consequently we can extend by continuity $\varphi(\vv{s})$ to $s_i = \mu^-$.
\end{itemize}

By induction we therefore obtain a continuous extension of $\varphi$. It remains to show that the order in which the dimensions are considered does not affect the result. For any permutation $\sigma$, let $\psi_\sigma$ be the continuous extension obtained by considering the players in the order $\sigma$. We wish to show that they are all equal. We can do this by proving the equality for any two permutations that differ only by switching two elements that are consecutive (according to the permutations). Let $\psi_1$ and $\psi_2$ be two permutations corresponding to this situation. Let $i$ and $j$ be the players that were permutated. We choose $\psi_1$ the case where $i$ is changed before $j$. Let $\vv{r}$ and $\vv{s}$ differ only in $i$ with $s_i = \mu^-$. Let $\vv{t}$ and $\vv{s}$ that only differ in $j$
with $t_j \in \{\mu^+,\mu^-\}$. Let $\vv{u}$ that only differs from $\vv{t}$ in $i$ and only differs from $\vv{r}$ in $j$. Let us start with $t_j=\mu^-$:
\begin{itemize}
    \item Suppose that $\psi_1(\vv{t}) > \mu^-$. Then by strategy-proofness for $r_i$ and $r_j$ less than $\psi_1(\vv{t})$ we have $\psi_2(\vv{r}) =\psi_1(\vv{r}) =\psi_1(\vv{t})$. By strategy-proofness it follows that $\psi_2(\vv{t}) =\psi_1(\vv{t})$. Conversely if $\psi_2(\vv{t}) > \mu^-$ then $\psi_2(\vv{t}) =\psi_1(\vv{t})$.
    \item The only other case is $\psi_2(\vv{t}) =\psi_1(\vv{t}) = \mu^-$.
\end{itemize}

Let us now consider $t_j = \mu^+$.
Without loss of generality we will assume that $r_i \leq \psi_1(\vv{r}) \leq r_j$.
\begin{itemize}
\item If $\psi_1(\vv{t}) \in \Lambda$. Then for $r_i <\psi_1(\vv{t})<r_j$ we have $\psi_1(\vv{t}) = \psi_1(\vv{r}) = \psi_2(\vv{u}) = \psi_2(\vv{t})$.
\item If $\psi_1(\vv{t}) = \mu^-$ then no matter the choice of $r_i$ we have $r_j > \psi_1(\vv{r})$; hence $\psi_2(\vv{u})= \psi_1(\vv{r})$. Therefore $\psi_2(\vv{t}) = \psi_1(\vv{s}) \leq \psi_1(\vv{t}) = \mu^-$.
\item If $\psi_1(\vv{t}) = \mu^+$, then no matter the $r_i$ we have $r_i < \psi_2(\vv{r})$; as such  $\psi_1(\vv{s}) = \psi_2(\vv{r})$. Consequently, $\psi_2(\vv{t}) \geq \psi_2(\vv{u}) = \psi_1(\vv{t}) = \mu^+$.
\end{itemize}

\end{proof}

\section{Proof of theorem \ref{MCarac}.}\label{proof of MCarac}

This section aims to prove the following: the voting function $\varphi$ is strategy-proof iff there exists a phantom function $\alpha : \Gamma \rightarrow \Lambda$ such that:

\label{initial_characterization}

\begin{equation}
\forall \vv{r}\in \Lambda^n; \varphi(\vv{r}) := \left\{ \begin{array}{ll}
		\alpha_{\vv{\mu}^-} & \mbox{ if } \forall j, r_j \leq \alpha_{\vv{\mu}^-} \\
        \alpha_{\theta(\vv{r},r_i)} & \mbox{ if } (i \in N)
         \mbox{ and } r_i = min \{ r_j | r_j \geq \alpha_{\theta(\vv{r},r_i)} \} \\
        r_i & \mbox{ if }  (i\in N) 
        \mbox{ and } \forall \epsilon > 0, \alpha_{\theta(\vv{r},r_i + \epsilon)} \leq r_i \leq \alpha_{\theta(\vv{r},r_i)}    
    \end{array}
    \right.
\end{equation}

Throughout the proofs,  $x_i$ often denotes the $i$th smallest element of the voting profile $\vv{r}$.

\begin{definition} For a given phantom function $\alpha$, and a given voting profile $\vv{r}$ we define
$\alpha_{\vv{r},i} := \alpha \circ \theta(\vv{r},x_i)$ where $x_i$ is the $i$th smallest element of ${(r_k)}_{k\in N}$ and we define $\alpha_{\vv{r},n+1} := \alpha_{\vv{\mu}^-}=\alpha(\mu^-,...,\mu^-)$.
\end{definition}

\begin{lemma}
\label{lemme sep}
Let us fix a phantom function $\alpha$. For all $\vv{r}$, there exists an $i$ such that one of the following holds

\begin{itemize}
\item $r_i = min \{ r_j | r_j \geq \alpha_{\theta(\vv{r},r_i)} \}$.
\item $\forall \epsilon>0, \alpha_{\theta(\vv{r},r_i + \epsilon)} \leq r_i \leq \alpha_{\theta(\vv{r},r_i)} $.
\item $\forall j \in N, r_j < \alpha_{\vv{\mu}^-}$.
 \end{itemize}
\end{lemma}

\begin{proof}
The sequence $(x_i)_{[|1,n|]}$ is increasing with $i$ and $\alpha_{\vv{r},i}$ is decreasing with $i$.

Either $x_n < \alpha_{\vv{\mu}^-}$, $\alpha_{\vv{\mu}^+} \leq x_1$ or $(x_i)_n$ and $\alpha_{\vv{r},i}$ must cross. Suppose that they cross, there is an $i$ such that $x_{i-1} \leq \alpha_{\vv{r},i} \leq x_i$ or $\alpha_{\vv{r},i+1} \leq x_i \leq \alpha_{\vv{r},i}$.

\begin{enumerate}
\item If $x_{i-1} < \alpha_{\vv{r},i} \leq x_i$ then $\alpha_{\theta(\vv{r},x_i)} = \alpha_{\vv{r},i} \leq x_i$.

Also, for $r_k < \alpha_{\vv{r},i}$ there is $m < i$ with $r_k = x_m$ thus
\[r_k = x_m \leq x_{i-1} < \alpha_{\vv{r},i} = \alpha_{\theta(\vv{r},x_i)}.\]
Therefore $r_k$ is not in $\{r_j | r_j \geq \alpha_{\theta(\vv{r},x_i)}\}$ 
and $x_i = min \{ r_j | r_j \geq \alpha_{\theta(\vv{r},x_i)}\}$.
\item If $\alpha_{\vv{r},i+1} \leq x_i \leq \alpha_{\vv{r},i}$ then by definition $\alpha_{\theta(\vv{r},x_{i+1})} \leq x_i \leq \alpha_{\theta(\vv{r},x_i)}$. There is no $r_j$ strictly between $x_i$ and $x_{i+1}$, so for all $\epsilon > 0$:
\[
\{ j | r_j \geq x_i + \epsilon \} \subseteq \{ j | r_j \geq x_{i+1} \} \]
\[
\theta(\vv{r}, x_i + \epsilon) \leq \theta(\vv{r}, x_{i+1}) \\
\]

Therefore by monotonicity of $\alpha$,  $\forall \epsilon > 0, \alpha_{\theta(\vv{r},x_i + \epsilon)} \leq \alpha_{\theta(\vv{r},x_{i+1})}\leq x_i \leq \alpha_{\theta(\vv{r},x_i)}$
\item If $\alpha_{\vv{\mu}^+} < x_1$, then $x_1 = min \{ r_j | r_j \geq \alpha_{\theta(\vv{r},x_1)}\}$.
\end{enumerate}

\end{proof}

\begin{lemma}
\label{main def}
The function $\varphi$, defined from a fixed $\alpha$ as:
\begin{equation*}
\label{main eq}
\forall \vv{r}; \varphi(\vv{r}) := \left\{ \begin{array}{ll}
		\alpha_{\vv{\mu}^-} & \mbox{ if } \forall j, r_j \leq \alpha_{\vv{\mu}^-} \\
        \alpha_{\theta(\vv{r},r_i)} & \mbox{ if } (i \in N)
         \mbox{ and } r_i = min \{ r_j | r_j \geq \alpha_{\theta(\vv{r},r_i)} \} \\
        r_i & \mbox{ if }  (i\in N) 
        \mbox{ and } \forall \epsilon, \alpha_{\theta(\vv{r},r_i + \epsilon)} \leq r_i \leq \alpha_{\theta(\vv{r},r_i)}    
    \end{array}
    \right.
\end{equation*}
is properly defined for all $\vv{r}$.
\end{lemma}

\begin{proof}
Given lemma \ref{lemme sep} there is always an $i$ that verifies one of the properties that define $\varphi$. We must show that if different $i$ satisfy the property this does not lead to an ambiguous definition.
\begin{itemize}
\item If the first and second property are verified then $\alpha_{\vv{r},i} \leq r_i \leq \alpha_{\vv{\mu}^-}$. By monotonicity of $\alpha$, $\alpha_{\vv{\mu}^-} = \alpha_{\vv{r}, n+1} \leq \alpha_{\vv{r},i}$, therefore $\alpha_{\vv{r},i} = \alpha_{\vv{\mu}^-}$. There is no contradiction.
\item If the first and third property are verified then $\forall \epsilon, \alpha_{\theta(\vv{r},r_i + \epsilon)} \leq r_i \leq \alpha_{\vv{\mu}^-}$. Therefore by monotonicity of $\alpha$, $\alpha_{\vv{\mu}^-} = r_i$. There is no contradiction.
\item If the second is verified for a voter $i$ and the third for a voter $k$ with $r_i \leq r_k$, then we have $r_i = \min \{ r_j | r_j \geq \alpha_{\theta(\vv{r},r_i)} \}$ and  $\forall \epsilon,\alpha_{\theta(\vv{r},r_k + \epsilon)} \leq r_k \leq \alpha_{\theta(\vv{r},r_k)}$. Since $r_i \leq r_k$, we have $\alpha_{\theta(\vv{r},r_k)} \leq \alpha_{\theta(\vv{r},r_i)}$. Hence:
\[r_i \leq r_k \leq \alpha_{\theta(\vv{r},r_k)} \leq \alpha_{\theta(\vv{r},r_i)} \leq r_i.\]
This proves that $\alpha_{\theta(\vv{r},r_i)} = r_k$. This case does not lead to an ambiguous definition.
\item If the second is verified for a voter $i$ and the third for a voter $k$ with $r_k < r_i$, then we have $r_i = min \{ r_j | r_j \geq \alpha_{\theta(\vv{r},r_i)} \}$ and  $\forall \epsilon, \alpha_{\theta(\vv{r},r_k + \epsilon)} \leq r_k \leq \alpha_{\theta(\vv{r},r_k)}$.
Therefore by monotonicity of $\alpha$, for $\epsilon>0$ small enough that $r_i > r_k + \epsilon$:
\[\theta(\vv{r}, r_i) \leq \theta(\vv{r}, r_k + \epsilon)\]
\[\alpha_{\theta(\vv{r},r_i)} \leq \alpha_{\theta(\vv{r},r_k + \epsilon)}\leq r_k < r_i.\]
This is absurd since $r_i = min \{ r_j | r_j \geq \alpha_{\theta(\vv{r},r_i)} \}$. Therefore this case never happens.
\item Suppose the second is true for two different voters $i$ and $k$. Therefore $r_i = \min \{ r_j | r_j \geq \alpha_{\theta(\vv{r},r_i)} \}$ and $r_k = \min\{ r_j | r_j \geq \alpha_{\theta(\vv{r},r_k)} \}$.

Without loss of generality we can suppose that $r_i \leq r_k$. Therefore by monotonicity of $\alpha$, $\alpha_{\theta(\vv{r},r_k)} \leq \alpha_{\theta(\vv{r},r_i)} \leq r_i$. Since $r_k$ is the min of $\{ r_j | r_j \geq \alpha_{\theta(\vv{r},r_k)} \}$  we have that $r_k \leq r_i$. As such $r_k = r_i$. Therefore $\alpha_{\theta(\vv{r},r_k)} = \alpha_{\theta(\vv{r},r_i)}$. This case does not raise an ambiguous definition.
\item Suppose the third is true for two different voters $i$ and $k$. Therefore for all $\epsilon>0$, $\alpha_{\theta(\vv{r},r_i + \epsilon)} \leq r_i \leq \alpha_{\theta(\vv{r},r_i)} $ and $\alpha_{\theta(\vv{r},r_k + \epsilon)} \leq r_k \leq \alpha_{\theta(\vv{r},r_k)}$.

Let us suppose that  $r_i \neq r_k$. Without loss of generality, we will use $r_i < r_k$. Choose $\epsilon$ so that $r_k > r_i + \epsilon$. Then by monotonicity of $\alpha$:

$$\alpha_{\theta(\vv{r},r_i + \epsilon)} \leq r_i < r_k \leq \alpha_{\theta(\vv{r},r_k)} \leq \alpha_{\theta(\vv{r},r_i + \epsilon)}$$

This implies that $r_i = r_k$, which is absurd. This case does not cause an ambiguous definition of $\varphi$.

\end{itemize}
We have considered all the cases. The definition is not ambiguous.
\end{proof}

\begin{lemma}
\label{main right}
If $\varphi$ verifies strategy-proofness then there is a phantom function $\alpha$ such that :

\begin{equation}
\label{equation a}
\forall \vv{r}; \varphi(\vv{r}) := \left\{ \begin{array}{ll}
		\alpha_{\vv{\mu}^-} & \mbox{ if } \forall j, r_j \leq \alpha_{\vv{\mu}^-} \\
        \alpha_{\theta(\vv{r},r_i)} & \mbox{ if } (i \in N)
         \mbox{ and } r_i = min \{ r_j | r_j \geq \alpha_{\theta(\vv{r},r_i)} \} \\
        r_i & \mbox{ if }  (i\in N) 
        \mbox{ and } \forall \epsilon, \alpha_{\theta(\vv{r},r_i + \epsilon)} \leq r_i \leq \alpha_{\theta(\vv{r},r_i)}    
    \end{array}
    \right.
\end{equation}
\end{lemma}

\begin{proof}
Let us use the phantom function such that 

    \[\alpha(X) := \varphi(X)\]
where we consider the Cauchy-extension of $\varphi$ if necessary. We will show for each case used in the equation \ref{equation a} we obtain the desired value of $\varphi(\vv{r})$. Lemma \ref{lemme sep} provides that we have studied all the cases and that therefore the proof is complete.
\begin{enumerate}
\item Case $ \forall j, r_j \leq \alpha_{\vv{\mu}^-}$: 

By definition of $\alpha$, $\mu^- \leq \varphi(\mu^-,\dots,\mu^-)$. If $\mu^- = \varphi(\mu^-,\dots,\mu^-)$, then all voters voted $\mu^-$. As such, $\varphi(\vv{r})=\alpha_{\vv{\mu}^-}$ Else $\mu^- < \varphi(\mu^-,\dots,\mu^-)$. Strategy-proofness on each dimension gives us that $\varphi(\vv{r}) =  \varphi(\mu^-,\dots,\mu^-) = \alpha_{\vv{\mu}^-}$.
\item  Case $\exists i, r_i = \min \{ r_j | r_j \geq \alpha_{\theta(\vv{r},r_i)}\}$:

Let $X = \theta(\vv{r},r_i)$. Suppose that $\varphi(\vv{r}) < \alpha_X$ (resp. $\alpha_X < \varphi(\vv{r})$). By strategy-proofness changing to $\mu^+$ (resp. to $\mu^-$) the ballot of a voter $j$ such that $X_j = \mu^+$ (resp. $X_j = \mu^-$) does not change the outcome. Therefore by weak responsiveness of $\varphi$ we have $\varphi(\vv{r}) < \alpha_X \leq \varphi(\vv{r})$ (resp. $\varphi(\vv{r}) \leq \alpha_X < \varphi(\vv{r})$). We have reached a contradiction.

Therefore if $\exists i, r_i = min \{ r_j | r_j \geq \alpha_{\theta(\vv{r},r_i)}\}$ then $\varphi(\vv{r}) = \alpha_X$.
\item Case $\forall \epsilon > 0, \alpha_{\theta(\vv{r},r_i + \epsilon)} \leq r_i \leq \alpha_{\theta(\vv{r},r_i)}$:

Suppose that $\varphi(\vv{r}) < r_i$ (resp. $r_i < \varphi(\vv{r})$). Let $X = \alpha_{\theta(\vv{r},r_i)}$ (resp. $X = \lim_{\epsilon>0, \epsilon \rightarrow 0}\theta(\vv{r},r_i + \epsilon) $). By strategy-proofness changing to $\mu^+$ (resp. to $\mu^-$) the ballot of a voter $j$ such that $X_i = \mu^+$ (resp. $X_i = \mu^-$) does not change the outcome.
    
 Therefore by weak responsiveness of $\varphi$ we have $\varphi(\vv{r}) < r_i \leq \alpha_X \leq \varphi(\vv{r})$ (resp. $\varphi(\vv{r}) \leq \alpha_X \leq r_i < \varphi(\vv{r})$).
 
We have reached a contradiction. Therefore if $\forall \epsilon>0, \alpha_{\theta(\vv{r},r_i+\epsilon)} \leq r_i \leq \alpha_{\theta(\vv{r},r_i)}$, then $r_i = \varphi(\vv{r})$.
\end{enumerate}

\end{proof}

\begin{lemma}
\label{main left}
If there is a phantom function $\alpha$ such that :

\begin{equation*}
\forall \vv{r}; \varphi(\vv{r}) := \left\{ \begin{array}{ll}
		\alpha_{\vv{\mu}^-} & \mbox{ if } \forall j, r_j \leq \alpha_{\vv{\mu}^-} \\
        \alpha_{\theta(\vv{r},r_k)} & \mbox{ if } (k \in N)
         \mbox{ and } r_k = min \{ r_j | r_j \geq \alpha_{\theta(\vv{r},r_k)} \} \\
        r_k & \mbox{ if }  (k\in N) 
        \mbox{ and } \forall \epsilon, \alpha_{\theta(\vv{r},r_k + \epsilon)} \leq r_k \leq \alpha_{\theta(\vv{r},r_k)}    
    \end{array}
    \right.
\end{equation*}
 Then $\varphi$ verifies strategy-proofness.
\end{lemma}

\begin{proof}
First we prove weak responsiveness then strategy-proofness.
\begin{itemize}
\item Weak responsiveness: This proof is by induction. Let $s$ that differs from $r$ only in dimension $i$ and $r_i < s_i$. 

Let $k$ be such that $\varphi(\vv{r}) \in \{\alpha_{\theta(\vv{r},r_k)}, r_k\}$ when $\varphi(\vv{r}) \neq \alpha_{\vv{\mu}^-}$.
\begin{enumerate}
\item Case $\varphi(\vv{r}) = \alpha_{\vv{\mu}^-}$:

$\alpha_{\vv{\mu}^-}$ is the minimum of $\varphi$ therefore:
\[ \varphi(\vv{r}) \leq \varphi(\vv{s}).\]
\item \label{fourth case transitivity}Case $s_i \leq \varphi(\vv{r})$:

We have $\varphi(\vv{r}) \leq r_k$, therefore $s_i \leq r_k$ and $i \neq k$.

If $s_i = r_k$ then $r_k \leq \alpha_{\theta(\vv{r},r_k)}$. We have $\forall \epsilon, \theta(\vv{s},s_k+ \epsilon)= \theta(\vv{r},r_k+\epsilon)$ and ${\theta(\vv{s},s_k)} \geq {\theta(\vv{r},r_k)}$. Therefore, $\forall \epsilon, \alpha_{\theta(\vv{s},s_k + \epsilon)} \leq s_k \leq \alpha_{\theta(\vv{s},s_k)}$.

Else $s_i < r_k$ implies ${\theta(\vv{s},s_k)} = {\theta(\vv{r},r_k)}$ and ${\theta(\vv{s},s_k+ \epsilon)}= {\theta(\vv{r},r_k+\epsilon)}$. Therefore, $\forall \epsilon, \alpha_{\theta(\vv{s},s_k + \epsilon)} \leq s_k \leq \alpha_{\theta(\vv{s},s_k)}$.

We can conclude that

\[ \varphi(\vv{r}) = \varphi(\vv{s}).\]

\item  \label{third case transitivity} Case $\varphi(\vv{r}) < r_i$:

If $k = i$, then $\varphi(\vv{r})= \alpha_{\theta(\vv{r},r_k)}$. Let $s_l = \min\{s_j | s_j \geq \varphi(\vv{r})\}$. Then ${\theta(\vv{s},s_l)} = {\theta(\vv{r},r_k)}$ and $s_l = \min\{s_j | s_j \geq \alpha_{\theta(\vv{r},s_l)}\}$. Therefore:

\[\varphi(\vv{r})  =\varphi(\vv{s}).\]

Else if $k \neq i$, ${\theta(\vv{s},s_k)} = {\theta(\vv{r},r_k)}$ and for any $\epsilon $ such that $0 <\epsilon < r_i -r_k$, we have ${\theta(\vv{s},s_k+ \epsilon)}= {\theta(\vv{r},r_k+\epsilon)}$. We can conclude that:

\[\varphi(\vv{r})  =\varphi(\vv{s}).\]

\item \label{second case transitivity} Case $\varphi(\vv{r}) = r_i$:

Let $X$ be defined by $X_j = \mu^+$ iff $j =i$ or $r_j > r_i$. We have $\theta(\vv{r},r_i + \epsilon) < X \leq \theta(\vv{r},r_i)$.

Let $x =\min{ \{r_j | r_j > r_i\}}. $
\begin{enumerate}
\item Subcase $\alpha_X < r_i$:

If there where no $j\neq i$ such that $r_j = r_i$ then $i=k$ and we would have $X = {\theta(\vv{r},r_i)}$. As such, $\alpha_{\theta(\vv{r},r_i)} < r_i = \varphi(\vv{r})$. This is absurd.

Therefore such a $j$ exists. As such, without loss of generality, we can consider that $i \neq k$. It follows that
$\theta(\vv{s},s_k) = \theta(\vv{r},r_i)$ and $\theta(\vv{s},s_k+ \epsilon) = \theta(\vv{r},r_k + \epsilon)$

\[\varphi(\vv{s}) = s_k =r_k = \varphi(\vv{r}).\]
\item \label{first case transitivity}
Subcase $r_i \leq \alpha_X$ and $s_i \leq x$:

We have $X = \theta(\vv{s},s_i)$ and ${\theta(\vv{s},s_i+ \epsilon)} \leq {\theta(\vv{r},r_i+\epsilon)}$.

$$\varphi(\vv{s}) = min(\alpha_X, s_i) \geq r_i = \varphi(\vv{r})$$

\item Subcase $r_i \leq \alpha_X$ and $s_i > x$:

Let $t$ differ from $r$ only in dimension $r_i$ with $t_i = x$. 

Show by induction that $\varphi(r) \leq \varphi(t)$ (\ref{first case transitivity}) and $\varphi(t) \leq \varphi(s)$ (\ref{third case transitivity}) or (\ref{second case transitivity})
Transitivity gives:

$$\varphi(r) \leq \varphi(s)$$

\end{enumerate}
\item Case $r_i < \varphi(\vv{r}) < s_i$:

Let $t$ differ from $r$ only in dimension $r_i$ with $t_i = \varphi(\vv{r})$. 

Show by induction that $\varphi(r) \leq \varphi(t)$ (\ref{fourth case transitivity}) and $\varphi(t) \leq \varphi(s)$ (\ref{second case transitivity}).
Transitivity gives:

$$\varphi(r) \leq \varphi(s)$$

\end{enumerate}

There are a finite number of values $r_k$ and $\alpha_X$ between $r_i$ and $s_i$. As such by inductive steps we have shown that $\varphi$ is weakly responsive.
\item Strategy-proofness : Let $s$ that differs from $r$ only in dimension $i$. Suppose that $r_i < \varphi(\vv{r})$:
\begin{itemize}
\item If $r_i < s_i$, then by weak responsiveness $\varphi(\vv{r}) \leq \varphi(\vv{s})$. 
\item If $\varphi(\vv{r}) = \alpha_{\vv{\mu}^-}$ and $s_i < r_i$ then $\varphi(\vv{r}) =\varphi(\vv{s}) = \alpha_{\vv{\mu}^-}$.
\item If $s_i < r_i < \varphi(\vv{r}) = \alpha_{\theta(\vv{r},r_k)}$ then $s_k = min\{ s_j | s_j \geq \alpha_{\theta(\vv{r},s_k)}\}$ therefore $\varphi(\vv{r}) = \varphi(\vv{s})$.
\item If $s_i < r_i < \varphi(\vv{r}) = r_k$ then $\theta(\vv{r},r_k) = \theta(\vv{s},s_k)$ and $\theta(\vv{r},r_k+ \epsilon) = \theta(\vv{s},s_k+ \epsilon)$. Therefore $\varphi(\vv{r}) = \varphi(\vv{s})$.
\end{itemize}

The proof for $r_i > \varphi(\vv{r})$ is symmetrical in every case except:
\begin{itemize}
    \item If $\alpha_{\theta(\vv{r},r_k)} <r_i <s_i$ then $r_k = \min{\{ r_j | r_j \geq \alpha_{\theta(\vv{r},r_k)}\}}$. Let $l$ be such that $s_l = \min{\{ s_j | s_j \geq \alpha_{\theta(\vv{r},r_k)}\}}$. We have $s_l = \min{\{ s_j | s_j \geq \alpha_{\theta(\vv{r},s_l)}\}}$. It follows that $\varphi(\vv{r}) = \varphi(\vv{s})$.
\end{itemize}

\end{itemize}
\end{proof}

\begin{corollary}[Phantom function characterization]
\label{main}
The voting function $\varphi$ is strategy-proof iff there exists a phantom function $\alpha : \Gamma \rightarrow \Lambda$ :

\begin{equation}
\forall \vv{r}; \varphi(\vv{r}) := \left\{ \begin{array}{ll}
		\alpha_{\vv{\mu}^-} & \mbox{ if } \forall j, r_j \leq \alpha_{\vv{\mu}^-} \\
        \alpha_{\theta(\vv{r},r_i)} & \mbox{ if } (i \in N)
         \mbox{ and } r_i = min \{ r_j | r_j \geq \alpha_{\theta(\vv{r},r_i)} \} \\
        r_i & \mbox{ if }  (i\in N) 
        \mbox{ and } \alpha_{\theta(\vv{r},r_i + \epsilon)} \leq r_i \leq \alpha_{\theta(\vv{r},r_i)}    
    \end{array}
    \right.
\end{equation}

Where $\epsilon$ is always chosen small enough for $\vv{r}$. 
The four cases given in the definition of the function are exhaustive.
\end{corollary}

\begin{proof}
The proof is immediate by using lemma \ref{main def}, \ref{main right} and \ref{main left}.
\end{proof}


\section{Missing proofs for the new characterizations}

\subsection{Proof of Theorem \ref{Moulin-Type Characterization: General Case}} \label{Proof of Theorem Median General}

A voting rule $\varphi$ is strategy-proof iff there exists a phantom function $\alpha$ (the same as in Theorem \ref{ICarac}) such that:

\begin{equation}
    \forall \vv{r}; \varphi(\vv{r}) := med(r_1,\dots,r_n,\alpha_{X_0(\vv{r})},\alpha_{X_1(\vv{r})},\dots,\alpha_{X_n(\vv{r})}).
\end{equation}

\begin{proof}

This a consequence of Theorem \ref{ICarac}.  Let $\varphi$ be a strategy-proof voting rule defined by a phantom function $\alpha$ as in Theorem \ref{ICarac}. Let $\mu : \Lambda^{N} \rightarrow \Lambda$ be defined as:

\begin{equation*}
    \forall \vv{r},\mu(\vv{r}) := med(r_1,\dots,r_n,\alpha_{X_0(\vv{r})},\alpha_{X_1(\vv{r})},\dots,\alpha_{X_n(\vv{r})}).
\end{equation*}

Since $\varphi$ is the unique strategy-proof voting rule defined by $\alpha$ we only need to prove that $\varphi = \mu$.

We will use the shorthand $X_k := X_k(\vv{r})$ since the context is clear (e.g. $\vv{r}$ will be fixed). Considering the two cases in the formula of Theorem \ref{ICarac}:

\begin{itemize}
    \item Case $\varphi(\vv{r}) = \alpha_X$. \\
    Let us first show that there is $k$ such that $\alpha_X = r_k$ or $\alpha_X = \alpha_{X_k}$.
    
   Suppose that $\forall j, \alpha_X \neq r_j$ then $\mu^+(X) = \{j | r_j > \alpha_X\}$. Therefore there is only one choice for $X_k$ where $k=\#\mu^+(X)$, that is $X_k = X$. Therefore $\alpha_{X_k} = \alpha_X$. Given our assumption we have therefore shown that there is $k$ such that $\alpha_X = r_k$ or $\alpha_X = \alpha_{X_k}$. In other words $\alpha_X$ is one of the arguments given to the median function. It now remains to show that $\alpha_X$ is selected by the median (e.g. is the output of $\mu$).
    
    If $\mu^+(X) = \{j | r_j \geq \alpha_X\}$ then for all $j \in \mu^+(X)$ we have $r_j \geq \alpha_X$. Also, for any $l$ such that $l \geq k$, we have that $\alpha_{X_l} \geq \alpha_X$. Therefore we have $n+1$ values in the median formula of $\mu$ that are greater of equal to $\alpha_X$. A symmetrical proof gives us $n+1$ values that are less or equal to $\alpha_X$.
    
    If $\mu^+(X) \subset \{j | r_j \geq \alpha_X\}$ then there is $r_k = \alpha_X$. For all $j \in \mu^+(X)$ we have $r_k \leq r_j$. For any $l$ such that $l \geq \#\{j | r_j \geq \alpha_X\}$, we have that $\alpha_{X_l} \geq r_k$. Therefore we have $n+1$ values in the median formula of $\mu$ that are greater of equal to $\alpha_X=r_k$. A symmetrical proof gives us $n+1$ values that are less or equal to $\alpha_X$. Consequently, $\alpha_X$ is the median of our set of values, that is: $\mu(\vv{r}) = \alpha_X$.
    
    \item Case $\varphi(\vv{r}) = r_i$. \\
    Let $X$ and $Y$ be such that $\alpha_X \leq r_i \leq \alpha_Y$ and $\mu^+(X) = \{ j | r_i < r_j \} \wedge \mu^-(Y) = \{ j | r_i > r_j \}$. For any $l \geq \#\mu^+(X)$ we have $\alpha_{X_l} \geq r_i$. Therefore we have ($r_i$ included) $n+1$ elements that are greater or equal to $r_i$. A symmetrical proof gives us that there are $n+1$ that are lesser or equal to $r_i$. Consequently, $r_i$ is the median of our set of values, that is: $\mu(\vv{r}) = r_i$.
    
\end{itemize}

Therefore $\varphi = \mu$.
\end{proof}


\section{Missing proofs for additional properties}

\subsection{Proof of Proposition \ref{prop ordinality}} \label{Proof prop ordinality}

For a strategy-proof voting rule $\varphi: \Lambda^{N} \rightarrow \Lambda$ the following are equivalent:\\
(1) The phantom function $\alpha$ verifies $\alpha(\Gamma) = \{\mu^-,\mu^+\}$\\
(2) $\varphi$ is strictly responsive.\\
And when moreover $\Lambda$ is a rich\footnote{$\Lambda$ is rich if for any $\alpha < \beta$ in $\Lambda$ there exists a $\gamma\in \Lambda$ such that $\alpha<\gamma < \beta$.}, (1) and (2) are equivalent to:\\
(3) $\varphi$ is ordinal and not constant.

\begin{proof}

$(2) \rightarrow (1):$ Suppose that $\varphi$ is strategy-proof and strictly responsive. If there exists $X\in \Gamma$ such that $\alpha_X \not \in \{\mu^-,\mu^+\}$, then set $\vv{r}$ and $\vv{s}$ such that if $j \in \mu^+(X)$ then $r_j = \alpha_X, s_j = \mu^+$ and if $j \in \mu^-(X)$ then $r_j = \mu^-, s_j = \alpha_X$. We have $\varphi(\vv{s}) = \varphi(\vv{r})$ which contradicts strict responsiveness. Therefore $\alpha(\Gamma) \subseteq \{\mu^-,\mu^+\}$. If $\#\alpha(\Gamma) = 1$ then the function is a constant and is therefore not strictly responsive. Consequently, $\alpha(\Gamma) = \{\mu^-,\mu^+\}$. 

$(1) \rightarrow (2):$ Suppose that $\varphi$ is strategy-proof and $\alpha(\Gamma) = \{\mu^-,\mu^+\}$. For any $\varphi(\vv{r})$ and $\varphi(\vv{s})$ such that for each $k$, $r_k < s_k$. Since all the phantoms are extreme we have: $i$ and $j$ such that $r_i = \varphi(\vv{r})$ and $s_j =\varphi(\vv{s})$. Let $\vv{t}$ be defined as for all $k$ if $r_k < r_i$ then $t_k = r_k$, else if $s_k = r_i$ then $t_k = \frac{r_k + s_k}{2}$, else $t_k = s_k$. We have $\varphi(\vv{t}) \in \{ t_k\}$ therefore by weak responsiveness, since $r_i \not \in \{ t_k\}$, we have $r_i < \varphi(\vv{t}) \leq s_j$

$(3) \rightarrow (1):$ Suppose that $\varphi$ is strategy-proof and ordinal. If there exists $X\in \Gamma$ such that $\alpha_X \not \in \{\mu^-,\mu^+\}$, let $\vv{r}$ be such that there are two alternatives $a < b < \alpha_X$ such that if $i \in \mu^+(X)$ then $r_i = b$ else $r_i = a$. Let $p_i$ be bijective and $\pi(a) < \alpha_X < \pi(b)$. Then:

\begin{align*}
 \alpha_X &= \varphi(\pi(r_1),\dots,\pi(r_n)) \\
    &= \pi \circ \varphi(\vv{r}) \\
    &= \pi(b) \\
    &> \alpha_X
\end{align*}

We have reached a contradiction, therefore $\alpha(\Gamma) = \{\mu^-,\mu^+\}.$

$(1) \rightarrow (3):$ Here we use the median representation in Theorem \ref{Moulin-Type Characterization: General Case}. Suppose that $\varphi$ is strategy-proof and $\alpha(\Gamma) = \{\mu^-,\mu^+\}$. For any strictly responsive and bijective $\pi$ and for any voting profile $\vv{r}$:

\begin{align*}
    \varphi(\pi(r_1),\dots,\pi(r_n)) &= med(\pi(r_1),\dots,\pi(r_n), \alpha_{X_0(\pi(\vv{r}))},\alpha_{X_1(\pi(\vv{r}))},\dots,\alpha_{X_n(\pi(\vv{r}))}) \\
    &= med(\pi(r_1),\dots,\pi(r_n), \alpha_{X_0(\vv{r})},\alpha_{X_1(\vv{r})},\dots,\alpha_{X_n(\vv{r})})  \\
    &= med(\pi(r_1),\dots,\pi(r_n), \pi(\alpha_{X_0(\vv{r})}),\pi(\alpha_{X_1(\vv{r})}),\dots,\pi(\alpha_{X_n(\vv{r})}))) \\
    &= \pi \circ med(r_1,\dots,r_n,\alpha_{X_0(\vv{r})},\alpha_{X_1(\vv{r})},\dots,\alpha_{X_n(\vv{r})}) \\
    &= \pi \circ \varphi(\vv{r})
\end{align*}

 Therefore $\varphi$ is ordinal.
\end{proof}


\subsection{Proof of Theorem \ref{theo participation}}\label{Proof of theo participation}

A strategy-proof voting rule $\varphi^* : \Lambda^* \rightarrow \Lambda$ verifies participation iff with the order $\mu^- < \emptyset < \mu^+$, its associated phantom function $\alpha$ is weakly increasing.

\begin{proof}
$\Rightarrow:$ Suppose $\mu^- < \emptyset < \mu^+$. We will prove by \emph{reductio ad absurdum}. Assume that $\alpha$ is not increasing. Therefore there is an elector $i$, such that for $X$ and $Y$ that only differ in $i$ we have $X_i < Y_i$ and $\alpha(X) > \alpha(Y)$. By definition of a phantom function either $X_i = \emptyset$ or $Y_i = \emptyset$. 

Suppose $X_i = \emptyset$ (therefore $Y_i = \mu^+$), then for $\vv{r} = Y$ we have that by removing his vote voter $i$ contradicts the participation property (e.g. participation).

A similar proof works for $Y_i = \emptyset$.

$\Leftarrow:$ Suppose that $\alpha$ is weakly increasing. Let $\varphi^*(\vv{r}) = a$ where elector $i$ did not cast a ballot. Let $\vv{s}$ be the voting profile that is identical to $\vv{r}$ except that $s_i$ is a ballot.

We will use now the curve characterisation:
\[\forall \vv{r}; \varphi^*(\vv{r}) := \mbox{sup } \left\{ y \in \Lambda | \alpha_{\theta(\vv{r},y)} \geq y \right\}.\]

\begin{itemize}
    \item If $s_i \leq a$ then $\alpha_{\theta(\vv{r},s_i)} \geq \alpha_{\theta(\vv{r},a)} \geq s_i$ therefore $\varphi(\vv{s}) \geq s_i$. We also have that    $\alpha_{\theta(\vv{r},a)} \geq \alpha_{\theta(\vv{s},a)}$ therefore  $\varphi^*(\vv{r}) \geq  \varphi^*(\vv{s})$.
    \item If $s_i \geq a$ then $\alpha_{\theta(\vv{r},s_i)} \leq \alpha_{\theta(\vv{r},a)} \leq s_i$ therefore $\varphi(\vv{s}) \geq s_i$. We also have that  $\alpha_{\theta(\vv{r},a)} \leq \alpha_{\theta(\vv{s},a)}$ therefore  $\varphi^*(\vv{r}) \leq  \varphi^*(\vv{s})$.
\end{itemize}
It follows that not submitting ones true value $s_i$ cannot be beneficial.
\end{proof}

\subsection{Proof of Proposition \ref{prop Participation and Winning Coalitions}}\label{Proof of prop Participation and Winning Coalitions} 

In a vote by issue, a strategy-proof voting function verifies participation iff when a elector $i$ decides to become a voter with ballot $x$ then for any property $H$ containing $x$, if $W_H$ was a winning coalition of $H$ for the initial set of voters then $W_H \cup \{i\}$ is a winning coalition for the new set of voters.

\begin{proof}
$\Rightarrow:$ Suppose that $\alpha$ is weakly increasing for the order $\mu^- < \emptyset < \mu^+$.
Let $V$ be a fixed set of voters such that $i \not \in V$.
For $H = \{y \geq a\}$, $W \subseteq V$ is a winning coalition iff $X \in \Gamma^*$ such  $\mu^+(X) = W$ verifies $\alpha_X \geq a$. Let $Y$ that differs from $X$ only in dimension $i$ with $Y_i = \mu^+$, then $\alpha_Y \geq a$. Therefore $\mu^+(Y)$ is a winning coalition.

A similar proof works for $H = \{y \leq a\}$.

$\Leftarrow:$ Suppose that when a electorate $i$ decides to become a voter with ballot $x$ then for any property $H$ containing $x$, if $W_H$ was a winning coalition of $H$ for the initial set of voters then $W_H \cup \{i\}$ is a winning coalition for the new set of voters.

Let us take $x = \alpha_X$ where $i$ is not a voter for $X$. Then $\mu^+(X) \cup \{i\}$ (resp. $\mu^-(x)$) is a winning coalition for $\{ y \geq x\}$ (resp. $\{ y \leq x\}$) therefore $\alpha_Y \geq \alpha_X$ (resp. $\alpha_Y \leq \alpha_X$) where $Y$ differs from $X$ only in dimension $i$ and $Y_i = \mu^+$ (resp. $Y_i = \mu^-$).

\end{proof}

\subsection{Proof of Theorem \ref{Curve and Consistency}} \label{Proof of Theorem Curve and Consistency}

A SP voting function $\varphi^*=(\varphi^n) : \Lambda^* \rightarrow \Lambda$ is anonymous and consistent iff there is an increasing function $g :[0,1] \rightarrow \Lambda$ (electorate size independent) and a constant $x \in \Lambda$ such that the phantom function $\alpha :\Gamma^*$ is defined as:

\[\alpha_X :=\left\{ \begin{array}{lll}
		g\left(\dfrac{\# \mu^+(X)}{\# N}\right) & \mbox{ if } & \# N \neq 0\\
		x & \mbox{ if } & \# N = 0
    \end{array}
    \right.\]
    
Furthermore the voting function verifies participation iff $x \in g([0,1])$.

\begin{proof}
$\Rightarrow:$ 
Let us first show the existence of $x$ and $g :[0,1] \rightarrow \Lambda$ such that forall $X \in \Gamma^*$ the equation holds.
$x = \alpha_{\vv{\emptyset}}$ therefore $x$ exists.
For any other $X$ if $q=\frac{\# \mu^+(X)}{\# \mu^-(X) + \# \mu^+(X)}$ 
we define $g(q)=\alpha_X$. Observe that this is well defined as by consistency and anonymity, we can duplicate and merge the electorate and so we must have $\alpha_X = \alpha_Y = g(q)$ whenever  $\frac{\# \mu^+(Y)}{\# \mu^-(Y) + \# \mu^+(Y)}=q$.

Now let us show that $g$ is increasing. For any $X$ and $Y$ such that $\frac{\# \mu^+(X)}{\# \mu^-(X) + \# \mu^+(X)} \leq \frac{\# \mu^+(Y)}{\# \mu^-(Y) + \# \mu^+(Y)}$ by consistency we can duplicate $X$ and $Y$ into $X'$ and $Y'$ that have the same number of voters. As such $\alpha_X \leq \alpha_Y$. It follows that $g$ is increasing over the set of rationals. Since the values over the irrationals do not matter we can complete the definition with an increasing $g$ without loss.

$\Leftarrow:$ 
For any $X \in \Gamma^*$ we have $\alpha_X = \alpha_{X \sqcup \vv{\emptyset}}$.

By using the barycentric weights we have that for any $X$ and $Y$ we have:
\[\frac{\# \mu^+(X)}{\# \mu^-(X) + \# \mu^+(X)} \leq \frac{\# \mu^+(X \sqcup Y)}{\# \mu^-(X \sqcup Y) + \# \mu^+(X \sqcup Y)} \leq \frac{\# \mu^+(Y)}{\# \mu^-(Y) + \# \mu^+(Y)}\]

Therefore since $g$ is increasing. $\alpha_X \leq \alpha_{X \sqcup Y} \leq \alpha_Y$. It follows that we verify consistency.

Finally we wish to show that if $x \in g([0,1])$ then our voting function verifies participation. For any $n > 0$ and $0 \leq k < n$ we have:
\[\frac{k}{n+1} < \frac{k}{n} < \frac{k+1}{n+1} \]
Therefore since $g$ is increasing $\alpha$ is increasing except maybe in $\alpha_{\vv{\emptyset}}$. Therefore the function verifies participation iff $x \in g([0,1])$. 
\end{proof}
`

\subsection{Proof of Theorem \ref{Theo Continuous Grading Curves}} \label{Proof of Theo Continuous Grading Curves}

A strategy-proof, homogeneous ($=$ consistent and anonymous) voting function $\varphi^* : \Lambda^* \rightarrow \Lambda$ is continuous with respect to new members iff its grading curve $g$ is continuous.

\begin{proof}

$\Rightarrow:$ We have $\forall \vv{r},\vv{s},\mbox{lim}_{n \rightarrow +\infty} \varphi(\overbrace{\vv{r} \sqcup \dots \sqcup \vv{r}}^n \sqcup \vv{s}) = \varphi(\vv{r})$. Therefore for any $X, Y \in \Gamma^*$ with at least one voter each:
\[ \mbox{lim}_{n \rightarrow +\infty} g\left(\frac{n \#  \mu^+(X) + \# \mu^+(Y)}{n (\# \mu^-(X) + \# \mu^+(X)) + \# \mu^-(Y) + \# \mu^+(Y)} \right) = g\left(\frac{\# \mu^+(X)}{\# \mu^-(X) + \# \mu^+(X)}\right)\]

Therefore since $g$ is increasing $g$ is continuous in all rational numbers. Therefore by monotonicity and density of the rationals within the real numbers we have that $g$ is continuous.

$\Leftarrow:$ Let $g$ be continuous. Then for all $X, Y$:
\[ \mbox{lim}_{n \rightarrow +\infty} g\left(\frac{n \#  \mu^+(X) + \# \mu^+(Y)}{n (\# \mu^-(X) + \# \mu^+(X)) + \# \mu^-(Y) + \# \mu^+(Y)} \right) = g\left(\frac{\# \mu^+(X)}{\# \mu^-(X) + \# \mu^+(X)}\right)\]
Therefore by continuity of $\varphi$.

\[\forall \vv{r},\vv{s},\mbox{lim}_{n \rightarrow +\infty} \varphi(\overbrace{\vv{r} \sqcup \dots \sqcup \vv{r}}^n \sqcup \vv{s}) = \varphi(\vv{r}).\]

\end{proof}

\section{Maximizing social welfare for weighted voters}

\subsection{Maximizing Ex-post Welfare  for Weighted Voters} \label{Proof Ex-Post Welfare}

\begin{definition}[Social welfare with for weighted voters]
Let $w_i$ be the weight of voter $i$. The social welfare for a given voting rule $\varphi : \Lambda^n \rightarrow \Lambda$ is defined as:
$$SW(\varphi,\vv{r}):= - \sum_i w_i \| \varphi(\vv{r}) - r_i \|_q$$
for a given norm $L_q$.
\end{definition}

For weighted inputs, the $L_1$-optimal voting rules are the weighted medians. The SP ones are therefore those that output the median of the two endpoints of the gap between the top 50-percentile and the bottom 50-percentile with some pre-determined real number. The $L_2$-optimal voting rule is the weighted mean $\varphi (\vv{r}) = \sum_i \frac{w_i}{\sum_j w_j} r_i$. This is therefore not a SP rule (except when the weights designate a dictator $i$: all $w_j=0$ except $w_i$).

\begin{theorem}
Let $p>1$, $n = \# N \geq 2$, $\Lambda = [m,M]$ and weights $w_i$ in $(\mathbf{R}^{+})^n$.  There is a unique voting rule that maximizes $SW$.  It is not SP unless the weights designate a dictator.
\end{theorem}

\begin{proof}
Fix $\vv{r}$ in $\Lambda^n$.  As a function of $r$, $|r - r_i|^p$ is strictly concave up with a minimum at $r_i$.  It is strictly decreasing on $(-\infty, r_i)$ and strictly increasing on $(r_i, \infty)$.

As long as at least one weight $w_i$ is greater than zero, $\sum_i w_i|r - r_i|^p$ is strictly concave up, strictly decreasing on $(-\infty, \min_i \{r_i\})$ and strictly increasing on $(\max_i \{r_i\}, \infty)$, so it will have a unique minimum which must be in $[\min_i \{r_i\},\max_i \{r_i\}]\subset [m,M]$.  The voting rule that minimizes $\sum_i w_i|r - r_i|^p$ for each $\vv{r}$ in $\Lambda^n$ is the unique voting rule that maximizes $SW$.

If the weights designate voter $j$ as a dictator, then this voting rule will always give $r_j$ as output, which is SP.

To show that if the weights do not designate a dictator then the $SW$-maximizing voting rule is not SP, suppose there are $w_i, w_j > 0$ with $i \ne j$.  Let $\varphi$ be the unique voting rule that maximizes $SW$ as described above.

For $\vv{s}$ with $s_k=m$ for all $k \ne i$, let $s = \varphi(\vv{s})$ be the outcome that maximizes $SW$.  Since we know $m \le s \le s_i$, it follows that:

\[\frac{d}{ds} \sum_k w_k |s - s_k|^p = 0\]

\[\frac{d}{ds} w_i (s_i - s)^p + \sum_{k\ne i} w_k (s - m)^p = 0\]

\[w_i (s_i - s)^{p-1} = \left(\sum_{k\ne i} w_k\right) (s - m)^{p-1}\]

\[\varphi(\vv{s}) = s = m + (s_i-m)\left(1 + \left(\frac{\sum_{k \ne i} w_k}{w_i}\right)^{\frac{1}{p-1}}\right)^{-1}\]

But $0 < \left(1 + \left(\frac{\sum_{k \ne i} w_k}{w_i}\right)^{\frac{1}{p-1}}\right)^{-1} < 1$, so $\varphi(\vv{s})$ increases as $s_i$ increases, but is always strictly less than $s_i$.  This proves that $\varphi$ is not SP.

\end{proof}

\subsection{Maximizing ex-ante welfare for weighted voters}\label{Proof Ex-Ante Welfare}




\begin{theorem}\label{norm-minimize}
Let $p$ be a probability distribution on $\mathbf{R}$ with support $\Lambda = [m,M]$.

If $G_p^q$ is strictly increasing, then $\alpha_X = g_p^q(\frac{\sum_i w_i X_i}{\sum w_i})$ generates the SP voting function which minimizes the $w_i$-weighted $L^q$-norm between inputs and output, integrated with the $L^q$-norm over all inputs (i.i.d. from $p$).  It is voter-sovereign.
\end{theorem}
\begin{proof}
Fix natural number $n$ and a probability distribution $p$ with compact support $[m,M]$.  We desire to find the SP voting function $f$ that minimizes

\[ E(f) = \int_m^M \cdots\int_m^M \sum_{i=1}^n w_i|x_i-f(\vv{x})|^q \prod_{i=1}^n p(x_i) dx_n\ldots dx_1\]

Since $f$ is an SP voting function, it is fully characterized by its $\alpha_X$ values as in lemma \ref{initial_characterization}.  Thus we can minimize $E(f)$ by optimizing each $\alpha_X$ value independently.  Fix $X$.

\[ \frac{\partial E}{\partial \alpha_X} = \sum_{i=1}^n w_i \int_m^M \cdots\int_m^M q |x_i-f(\vv{x})|^{q-1} \frac{\partial}{\partial \alpha_X}|x_i-f(\vv{x})| \prod_{j=1}^n p(x_j) dx_n\ldots dx_1\]

Let $k_1, k_2, \ldots, k_a$ be the indices of $X$ with $X_{k_i} = 1$ and $\hat{k}_1, \hat{k}_2, \ldots, \hat{k}_b$ be the indices of $X$ with $X_{\hat{k}_i} = 0$.  $\frac{\partial}{\partial \alpha_X} f$ will be 1 where $f(\vv{x}) = \alpha_X$ (and 0 elsewhere) which is exactly in the region where $
x_{\hat{k}_1}, x_{\hat{k}_2}, \ldots, x_{\hat{k}_b}
< \alpha_X <
x_{k_1}, x_{k_2}, \ldots, x_{k_a}
$.  Thus,

\[ \frac{1}{q} \frac{\partial E}{\partial \alpha_X}
=\]
\[
\left(
\sum_{i=1}^b w_{\hat{k}_i}
\underbrace{\int_{\alpha_X}^M \cdots\int_{\alpha_X}^M}_a
\underbrace{\int_m^{\alpha_X} \cdots\int_m^{\alpha_X}}_b
(\alpha_X - x_{\hat{k}_i})^{q-1}
\prod_{j=1}^n p(x_j)
dx_{\hat{k}_1}\ldots dx_{\hat{k}_b}
dx_{k_1}\ldots dx_{k_a}
\right.\]
\[\left.
-
\sum_{i=1}^a w_{k_i}
\underbrace{\int_{\alpha_X}^M \cdots\int_{\alpha_X}^M}_a
\underbrace{\int_m^{\alpha_X} \cdots\int_m^{\alpha_X}}_b
(x_{k_i}-\alpha_X)^{q-1}
\prod_{j=1}^n p(x_j)
dx_{\hat{k}_1}\ldots dx_{\hat{k}_b}
dx_{k_1}\ldots dx_{k_a}
\right)
\]

\[
= \left(
\sum_{i=1}^b w_{\hat{k}_i}
\left(\int_{\alpha_X}^M p(x) dx\right)^{a}
\left(\int_m^{\alpha_X} p(x) dx\right)^{b-1}
\int_m^{\alpha_X}
(\alpha_X - x_{\hat{k}_i})^{q-1} p(x_{\hat{k}_i})
dx_{\hat{k}_i}
\right.\]
\[\left.
-
\sum_{i=1}^a w_{k_i}
\left(\int_{\alpha_X}^M p(x) dx\right)^{a-1}
\left(\int_m^{\alpha_X} p(x) dx\right)^{b}
\int_{\alpha_X}^M
(x_{k_i}-\alpha_X)^{q-1} p(x_{k_i})
dx_{k_i}
\right)
\]

It follows that
\[ \left(\frac{1}{q \left(\int_{\alpha_X}^M p(x) dx\right)^{a}\left(\int_m^{\alpha_X} p(x) dx\right)^{b}
}\right) \frac{\partial E}{\partial \alpha_X}\]
\[=
\left(\sum_{i=1}^b w_{\hat{k}_i}\right)
\frac{\int_m^{\alpha_X} (\alpha_X - x)^{q-1} p(x) dx}
{\int_m^{\alpha_X} p(x) dx}
-
\left(\sum_{i=1}^a w_{k_i}\right)
\frac{\int_{\alpha_X}^M (x-\alpha_X)^{q-1} p(x) dx}
{\int_{\alpha_X}^M p(x) dx}
\]

This is negative iff:
\[
\frac{\sum_{i=1}^b w_{\hat{k}_i}}{\sum_{i=1}^a w_{k_i}}
<
\frac{\frac{\int_{\alpha_X}^M (x-\alpha_X)^{q-1} p(x) dx}
{\int_{\alpha_X}^M p(x) dx}}
{\frac{\int_m^{\alpha_X} (\alpha_X - x)^{q-1} p(x) dx}
{\int_m^{\alpha_X} p(x) dx}}
\]

\[
\frac{\sum_i w_i X_i}{\sum_i w_i}
>
\left(1 + \frac{\frac{\int_{\alpha_X}^M (x-\alpha_X)^{q-1} p(x) dx}
{\int_{\alpha_X}^M p(x) dx}}
{\frac{\int_m^{\alpha_X} (\alpha_X - x)^{q-1} p(x) dx}
{\int_m^{\alpha_X} p(x) dx}}
\right) ^ {-1}
=
G_p^q(\alpha_X)
\]

\[\alpha_X < g_p^q\left(\frac{\sum_i w_i X_i}{\sum_i w_i}\right)\]

Thus if $G_p^q$ is strictly monotone, $g_p^q(\frac{\sum_i w_i X_i}{\sum_i w_i})$ is the optimal $\alpha_X$ with respect to the $L^q$ norms.

Further, $\alpha_{\vv{0}} = m$ and $\alpha_{\vv{1}} = M$ so this voting function is voter-sovereign.
\end{proof}

\begin{corollary}
Let $p$ be a probability distribution on $\mathbf{R}$ with support $\Lambda = [m,M]$.

If $G_p^q$ is strictly increasing, then $\alpha_X = g_p^q(\frac{\sum_i X_i}{\# N})$ generates the anonymous SP voting function which minimizes the $L^q$-norm between inputs and output, integrated with the $L^q$-norm over all inputs (i.i.d. from $p$).  It is voter-sovereign and consistent.
\end{corollary}

\begin{proof}
Immediate from theorem \ref{norm-minimize} using equal weights.
\end{proof}

If $G_p^q$ is not strictly monotone, it may still be possible to minimize $E(f)$ for each $\alpha_X$ and invert to find a grading curve. In order to find a voting function that is consistent, we need to ensure that our inverse function is increasing. If not, it is possible that it would generate a family of grading curves that is not consistent.

\subsection{Maximizing worst case ex-ante welfare for weighted voters} \label{Proof MinMax Welfare}
\begin{theorem}
For $q\ge 1$ and a compact interval $[m,M]$, for all $n$,

\[\alpha_X = m + (M - m)\left(1 + \left(\frac{\sum_i w_i}{\sum_i w_i X_i}-1\right)^{\frac{1}{q-1}}\right)^{-1}\]

generates an SP voting function which minimizes the worst case (minimax) $w_i$-weighted $L^q$-norm between inputs and outputs.  It is voter-sovereign.
\end{theorem}

\begin{proof}

Fix $q>1$, compact interval $[m,M]$, natural number $n$, and voter weights $w_i$.  We desire to find the SP voting function $f$ that minimizes

\[ E(f) = \max_{\vv{x} \in [m,M]^n} \sum_{i=1}^n w_i|x_i-f(\vv{x})|^q\]

Since $f$ is an SP voting function, it is fully characterized by its $\alpha_X$ values as in lemma \ref{initial_characterization}.

Furthermore, $\sum_{i=1}^n w_i|x_i-f(\vv{x})|^q$ does not achieve its maximum in a region where $f$ returns a non-$\alpha$ value.  (One can always increase the $L^q$-norm by moving to the boundary of that region, moving into a region where $f$ returns one of the $\alpha$ values, and then moving all the input votes to the extremes.)

Fix $X$.  Let $k_1, k_2, \ldots, k_a$ be the indices of $X$ with $X_{k_i} = 1$ and $\hat{k}_1, \hat{k}_2, \ldots, \hat{k}_b$ be the indices of $X$ with $X_{\hat{k}_i} = 0$.  $f(\vv{x}) = \alpha_X$ in the region where $
x_{\hat{k}_1}, x_{\hat{k}_2}, \ldots, x_{\hat{k}_b}
< \alpha_X <
x_{k_1}, x_{k_2}, \ldots, x_{k_a}
$.  And in this region, we maximize $\sum_{i=1}^n w_i|x_i-f(\vv{x})|^q$ by setting $x_{\hat{k}_1} = x_{\hat{k}_2} = \ldots = x_{\hat{k}_b} = m$ and
$x_{k_1} = x_{k_2} = \ldots = x_{k_a} = M$.

We then find the $\alpha_X$ that minimizes this expression
\[\sum_{i=1}^a w_{k_i}(M-\alpha_X)^q + \sum_{i=1}^b w_{\hat{k}_i}(\alpha_X-m)^q\]
by setting the derivative equal to 0.

\[\frac{d}{d\alpha_X} \sum_{i=1}^a w_{k_i}(M-\alpha_X)^q + \sum_{i=1}^b w_{\hat{k}_i}(\alpha_X-m)^q = 0\]

\[\left(\sum_{i=1}^a w_{k_i}\right)(M-\alpha_X)^{q-1} =
\left(\sum_{i=1}^b w_{\hat{k}_i}\right)(\alpha_X-m)^{q-1}\]

\[\left(\frac{\sum_i w_i}{\sum_i w_i X_i}-1\right)^{\frac{1}{q-1}} =
\frac{M-\alpha_X}{\alpha_X-m}\]

\[\alpha_X = m + (M - m)\left(1 + \left(\frac{\sum_i w_i}{\sum_i w_i X_i}-1\right)^{\frac{1}{q-1}}\right)^{-1}\]
\end{proof}

We note that, when seeking to minimize the maximum $L^q$-norm, it is sufficient but not always necessary to optimize every single $\alpha_X$ value as in the theorem above.  Some $\alpha_X$ values may have considerable leeway and can stray somewhat from their optimal values without having any negative effect on the minimax.

\section{Algorithms and complexity}

\begin{algorithm}
\DontPrintSemicolon 
\KwIn{The set of votes $r_1,\dots,r_n$}
\KwOut{The value of $\varphi(\vv{r})$}

\For{$X \in \Gamma$}{
\If{$\mu^+(X) = \{r_j \geq \alpha_X\}$}{
\Return{$\alpha_X$}}}

\For{$i \in N$}{
	$X := \vv{\mu}^-$;
	
	$Y := \vv{\mu}^-$;

	\For{$j\in N$}{
	\If{$r_j > r_i$}{
	$Y_j := \mu^+$;
	
	$X_j :=\mu^+$}
	\ElseIf{$r_j = r_i$}{
	$Y_j = \mu^+$}}
	\If{$\alpha_X \leq r_i \leq \alpha_Y$}{
	\Return{$r_i$}}
}

\caption{$\varphi$ with Theorem \ref{ICarac} characterization}
\label{algo:main}
\end{algorithm}

\begin{lemma}
The complexity for Algorithm \ref{algo:main} is $O(2^n (n+f(n)))$.
\end{lemma}

\begin{proof}
In the first loop we iterate all $2^n$ elements of $\Gamma$. Therein, we need to calculate $\alpha_X$ and to check if $X =  \{r_j \geq \alpha_X\}$ hence $O(2^n (n+f(n)))$.

In the second main loop we test each voter $i$, in order to create $X$ and $Y$ we need to consider all voters $j$. Calculating $\alpha_X$ and $\alpha_Y$ take $O(f(n))$. It follows that the second loop has complexity $O(n^2 + nf(n))$. 

Therefore the complete complexity is $O(2^n (n+f(n)))$.
\end{proof}

\begin{algorithm}
\DontPrintSemicolon 
\KwIn{The set of votes $r_1,\dots,r_n$}
\KwOut{The value of $\varphi(\vv{r})$}
$l = [\alpha_{\vv{\mu}^-}]$;

\For{$i \in N$}{
$v_i = (r_i,i)$;\\
$l := add(r_i,l)$;
}
$ll:=$ Sort $\{v_i\}$ according to natural decreasing order on $r_i$;

$X := \vv{\mu}^-$;

\While{$ll \neq \emptyset$}{
$v_i = pop(ll)$;\\
$X_i \leftarrow \mu^+$;\\
$l:=add(\alpha(X),l)$;
}
\Return $med(l)$
\caption{$\varphi$ with the Theorem \ref{Moulin-Type Characterization: General Case} characterization}
\label{algo:moulin}
\end{algorithm}

\begin{lemma}
The complexity for Algorithm \ref{algo:moulin} is $O(n(\log(n)+f(n)))$.
\end{lemma}

\begin{proof}
The first loop takes complexity $n$. Then we need $O(n\log n)$ to sort the elements of $ll$. Then $O(nf(n))$ to complete our list of $n+1$ elements. And finally $O(n\log(n))$ to determine the final median value. 

It follows that the complexity is $O(n(\log(n)+f(n)))$.
\end{proof}

\begin{algorithm}
\DontPrintSemicolon 
\KwIn{The set of votes $r_1,\dots,r_n$}
\KwOut{The value of $\varphi(\vv{r})$}

\If{$max\{r_i\} \leq \alpha_{\vv{\mu}^-}$}{
\Return $\alpha_{\vv{\mu}^-}$.
}

$ll:=$ Sort $\{(r_i,i)\}$ according to natural decreasing order on $r_i$;

$X := \vv{\mu}^-$

$(r_0, i_0) := ll[0]$

$X_{i_0} \leftarrow \mu^+$

\While{$length~ll > 1$}{
$p = \lfloor\frac{length(ll)}{2}\rfloor$;

$(x, i) = ll[p]$;

$Y := X$

\For{$(r_j,j) \in ll[1, p]$}{
$Y_j \leftarrow \mu^+$

}

\If{$\alpha(Y) < x$ }{

$ll := ll[0,p-1]$;

}
\Else{
$ll := ll[p,length(ll)-1]$

$X \leftarrow Y$

}
}

$(a, i) = ll[0]$;

\If{$\alpha(X) < a$}{
\Return $\alpha_x$
}

\Else{
\Return $a$
 }
\caption{$\varphi$ with Theorem \ref{theo curve} characterization}
\label{algo:jennings}
\end{algorithm}

\begin{lemma}
The complexity for Algorithm \ref{algo:jennings} is $O((n+f(n))\log(n))$.
\end{lemma}

\begin{proof}
We need $O(n\log n)$ to create and sort $ll$.
Then we use a ($\log n$) binary search with inner loop complexity $O(n + f(n))$ to find the final value. As such the final complexity is $O((n+f(n))\log(n))$.
\end{proof}

\begin{lemma}
The Moulin minmax characterisation provides the solution with a complexity of $O(2^n (n+f(n)))$. 
\end{lemma}

\begin{proof}
For each $S \in 2^N$ we calculate $\beta_S$ and then find the minimum of $\beta_S$ and $\{r_i :i \in S\}$ in linear time $n$.
\end{proof}

\end{document}